\documentclass[apj,onecolumn]{emulateapj}
\input epsf.tex
\begin{document}
\title{Testing In Situ Assembly with the Kepler Planet Candidate Sample} 
\author{Brad M. S. Hansen\altaffilmark{1} \& Norm Murray\altaffilmark{2,3}
}
\altaffiltext{1}{Department of Physics \& Astronomy and Institute of Geophysics \& Planetary Physics, University of California Los Angeles, Los Angeles, CA 90095, hansen@astro.ucla.edu}
\altaffiltext{2}{Canadian Institute for Theoretical Astrophysics, 60
St.~George Street, University of Toronto, Toronto, ON M5S 3H8, Canada;murray@cita.utoronto.ca}
\altaffiltext{3}{Canada Research Chair in Astrophysics}


\lefthead{Hansen \& Murray}
\righthead{Kepler Rocks}

\begin{abstract}

We present a Monte Carlo model for the structure of low mass (total mass $< 25 M_{\oplus}$) planetary systems that form by the in situ gravitational
assembly of planetary embryos into final planets. Our model includes distributions of mass, eccentricity, inclination and period spacing that
are based on the simulation of a disk of $20 M_{\oplus}$, forming planets around a solar mass star, and assuming a power law surface density distribution
$\propto a^{-1.5}$.

The output of the Monte Carlo model is then subjected to the selection effects that mimic the observations of a transiting planet search such
as that performed by the Kepler satellite. The resulting comparison of the output to the properties of the observed sample yields an encouraging
agreement in terms of the relative frequencies of multiple planet systems and the distribution of the mutual inclinations, when moderate
tidal circularisation is taken into account. The broad features of the
period distribution and radius distribution can also be matched within this framework, although the model underpredicts the distribution of
small period ratios. This likely indicates that some dissipation is still required in the formation process.

The most striking deviation between model and observations is in the ratio of single to multiple systems, in that there are roughly 50\% more
single planet candidates observed than are produced in any model population. This suggests that some systems must suffer additional attrition
to reduce the number of planets or increase the range of inclinations.

\end{abstract}


\section{Introduction}

The search for planets around stars other than the Sun has revealed a great diversity of both planet mass
and location. Jovian mass planets have been discovered both in very short period orbits (Mayor \& Queloz 1995, Butler et al. 1997),
 at distances more characteristic of the planets in our own solar system (Butler \& Marcy 1996; Boisse et al. 2012), at great
distances from their host stars (Udalski et al. 2005; Gaudi et al. 2008; Kalas et al. 2008; Marois et al. 2008, 2010) and potentially even as
free floating objects (Sumi et al. 2011). Lower mass planets are harder to detect, but planets in
the Uranus/Neptune mass range ($10 M_{\oplus}$--$100 M_{\oplus}$) have also been found with a range of semimajor axes, both close
in (Howard et al. 2010; Mayor et al. 2011; Borucki et al. 2010; Holman et al. 2010; Lissauer et al. 2011a) and at 
large separations (Gould et al. 2006, 2010; Sumi et al. 2010).  Furthermore, multiple detection techniques are now
able to probe below $10 M_{\oplus}$ (Wolszczan \& Frail 1992; McArthur et al. 2004; Santos et al. 2004;
 Rivera et al. 2005; Beaulieu et al. 2006; Lovis et al. 2006;  Charbonneau et al. 2009; Queloz et al. 2009; Mayor et al. 2009; Batalha et al. 2011), potentially delving into the regime of true terrestrial planet formation.
Indeed, the number of planet candidates emerging from the Kepler satellite data suggests that this class may soon dominate
the demographics of known extrasolar planets (Borucki et al. 2011).

The origin of these different classes of planet is still a matter of active discussion. The presence of so-called hot Jupiters
close to their parent stars was initially held to be securely described within the context of planetary migration (Lin, Bodenheimer \& Richardson 1996),
in which giant planets form at distances $\sim$ several~AU, and migrate inwards due to torques from the gaseous protoplanetary
disk (Goldreich \& Tremaine 1980; Ward 1997). The discovery of significant misalignments between the planetary orbital angular momenta and the stellar star host spin
(Winn et al. 2010 and historical references therein) has cast some doubt on the applicability of this model to at least some of the observed systems, reviving interest
in such mechanisms as planetary scattering (Rasio \& Ford 1996; Nagasawa \& Ida 2011) or Kozai migration (Wu \& Murray 2003; Fabrycky \& Tremaine 2007; Naoz et al. 2011). Another significant problem
for the migration paradigm is the fact that both radial velocity and transit methods find Neptune mass planets in a part of
parameter space that such models predicted to be devoid of planets (Ida \& Lin 2008). This does not necessarily imply that
migration does not occur, but it does indicate that current models are either not universally applicable or are missing one or more crucial
elements.

Such advances have spurred new interest in the origin of low mass planets around other stars, and generated alternative proposals.
It is possible that such planets did migrate inwards but were subject to additional perturbations, either from mutual gravitational
interactions (Terquem \& Papaloizou 2007; Ida \& Lin 2010) or some
 other, stochastic, forcing (Rein 2012), in
order to match the observed period ratio distribution in multiple systems. Alternatively, it has been proposed that intense stellar
irradiation can evaporate the envelopes of hot Jupiter planets, leaving behind only the Neptune-mass stellar core (Baraffe et al. 2004).
However, current estimates of the evaporation rates for known Jupiter planets (Ehrenreich \& Desert 2011) suggest that these do not 
compose a viable progenitor population. 

Instead, Hansen \& Murray (2011) propose that this population is the result of in situ assembly
of rocky planets, which may capture substantial, but not dominant, gaseous envelopes if they grow to sufficient size before the gaseous
disk disperses. This proposal has the advantage that it naturally produces multiple planet systems, as are observed, and furthermore
matches the statistical properties of the observed multiple planet systems.

The original intent of the Hansen \& Murray (2011) proposal was to match the planets observed in the radial velocity surveys descibed in
 Howard et al. (2010) and Mayor et al. (2011), which represent the tip of the iceberg of the low mass planet sample, by virtue of the
detection limits inherent in the RV method. However, the Kepler mission (Borucki et al. 2011) has begun to expose more of the iceberg, suggesting
that there are many systems with lower mass planets, offering the possibility of further testing the in situ assembly model. In particular,
these lower mass planets most likely did not substantially increase their masses by accreting gas from the nebula and so should represent a more straightforward test of
the model, without uncertainties in the treatment of gas accretion. Therefore, in
 this paper, we perform a quantitative comparison of in situ assembly models with a
planetary candidate sample culled from the Kepler satellite data, under the assumption that these are primarily rocky planets and that
their organization reflects the conditions that emerge directly from the gravitational assembly of the planetary system. 

In \S~\ref{Sim} we describe our model for the original rocky protoplanetary disk, and its assembly into the final planetary configuration.
We also develop statistical descriptions for the distribution of various important quantities, such as planetary masses and mutual
inclinations which are then used to simulate the distribution of a large population of
planetary systems in a Monte Carlo fashion, and to quantify their observability when studied in
the manner of a search for transits. These are then compared to a well-defined subset of the Kepler candidates
in \S~\ref{Obs} and the implications are discussed in \S~\ref{Discuss}.

\section{Planetary Assembly Simulations}
\label{Sim}

Hansen \& Murray (2011), hereafter HM11, simulated the assembly of planets for a variety of rocky disks, with masses ranging up $\sim 100 M_{\oplus}$
within 1~AU, and accounting for both purely rocky systems and for the accretion of substantial amounts of gas from the surrounding nebula.
This was dictated by a desire to match systems detected in radial velocity surveys, whose masses suggest rock-to-gas ratios of order unity
(and whose existence is born out by a handful of transiting systems, such as HD149026b, HAT-P-11 and Kepler-9b,c). However, statistical
analyses of the Kepler sample (e.g. Howard et al. 2012, Youdin 2011) suggest that this population contains a substantial contribution from
bodies whose mass is predominantly rocky, even if their radii are inflated by a minority component (by mass) of gas. Therefore, this new
sample, by virtue of its size and different detection technique, offers a new test of planetary origin models. 

Our model for the origin of these planetary systems is that the final assembly stage from planetary embryos to final planets occurred
 in situ, with the masses and separations of the final planets determined by the gravitational interactions between the assembling bodies. Whether
the planetary embryos are formed from rocky material that condensed out of the gas in situ or migrated inwards as small bodies is not
specified, and may be a function of total disk mass. In HM11, the disk masses assumed were in excess of 30$M_{\oplus}$ and most likely
did require an enhancement over the heavy element inventory of the original local gas disk. However, for lower mass disks, such inward
migration may not be necessary.

Our model for the origin of the observed Kepler systems has a surface density profile $\Sigma \sim a^{-1.5}$, normalised to contain
$20 M_{\oplus}$ of rocky material interior to 1~AU, around a 1$M_{\odot}$ star. This is chosen to be a lower mass version of the density profile
that best fit the observations in HM11, and the overall mass normalisation was chosen to be the maximum that would not result in many of the
assembling protoplanets crossing the gas capture threshold defined in HM11 within 1~Myr. Interestingly, these criteria yield almost exactly the
same disk as has been recently derived by Chiang \& Laughlin (2012), using an analogue of the Minimum Mass Solar Nebula argument, but applied
to the observed Kepler candidate sample. This agreement suggests an encouraging self-consistency to our adoption of this as the underlying basis
for forming a Kepler sample dominated by rocky planets.

Our initial conditions are chosen using the formalism of Kokubo \& Ida (1998) to assign initial masses and semi-major axes for the protoplanetary
embryos at the end of the oligarchic accumulation stage, and are integrated forwards using the Mercury integrator (Chambers 1999) on the UCLA Hoffman2 Shared Computing Cluster. We perform 
100~realisations of this model.
 The inner edge of the original disk is taken to be at 0.05~AU, and so the time step for the integrations is
only 12~hours, in order to resolve the innermost orbits. The integrations are run for 10~Myr.
This proves sufficient to achieve dynamical stability, as evidenced both by inspection of the individual 
evolutionary histories and the integration of a subsample of 20 to ages of 100~Myr, without further significant change in the overall configuration.
This assembly time scale is shorter than has been found for the solar system terrestrial planets (e.g. Chambers \& Wetherill 1998; Chambers 2001) because of the higher mass and more compact nature of
the disk.

We then examine the ensemble of surviving systems to characterise various properties of the resulting population.
The parameters of the final systems are shown in Table~\ref{OutTab}, including a variety of statistical measures collected together by Chambers (2001)
for the purposes of characterising simulations of late stage planet assembly in the Solar System. Not surprisingly, the simulations here are both more closely packed
(because of the larger mass in the disks relative to the solar system case studied by Chambers) and have the mass distributed over a greater range of radii
(because we consider disks extending down to 0.05~AU). Furthermore, many of these measures are based on the distribution of mass, a quantity only
indirectly accessible for the bulk of the Kepler sample. As such, we will use our simulation results to build a Monte Carlo model to perform a more
direct comparison with the observables of a Kepler-style experiment.

\subsection{Multiplicity}

One of the most fundamental questions one can ask about the Kepler data set is how many planets a given system is likely to hold. Kepler has
detected 361 stars (Fabrycky et al. 2012) which exhibit multiple transiting planets, with the record for the largest number of transiting planets around a single star currently held by Kepler-11 (Lissauer et al. 2011a)
at six. Of course, the mutual inclinations of a set of planets orbiting a star will dictate what fraction transit, and therefore the transiting
planets represent only a subsample of the true underlying system. 

\begin{figure}
\centerline{\includegraphics[width=0.52\textwidth]{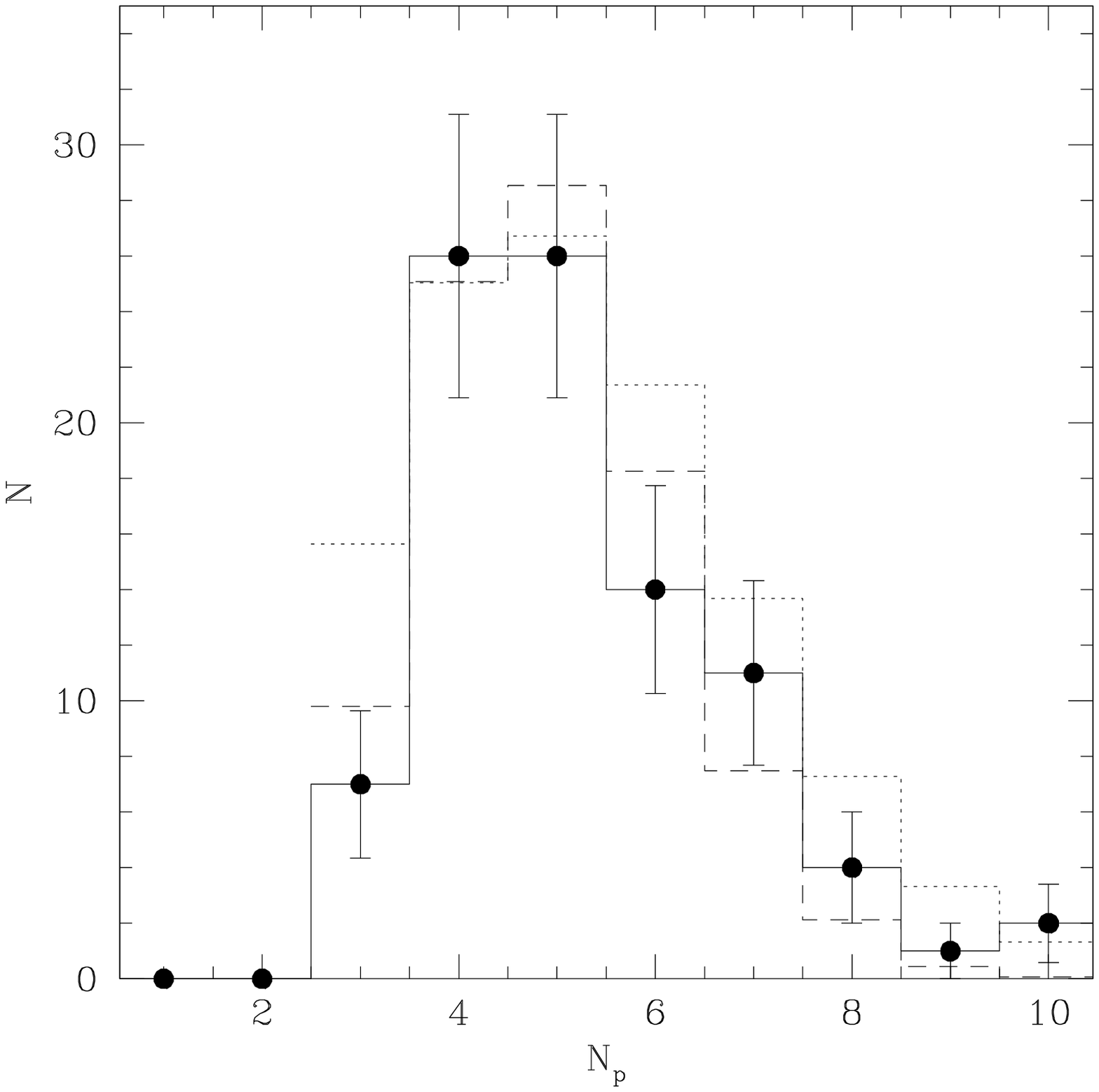},\includegraphics[width=0.52\textwidth]{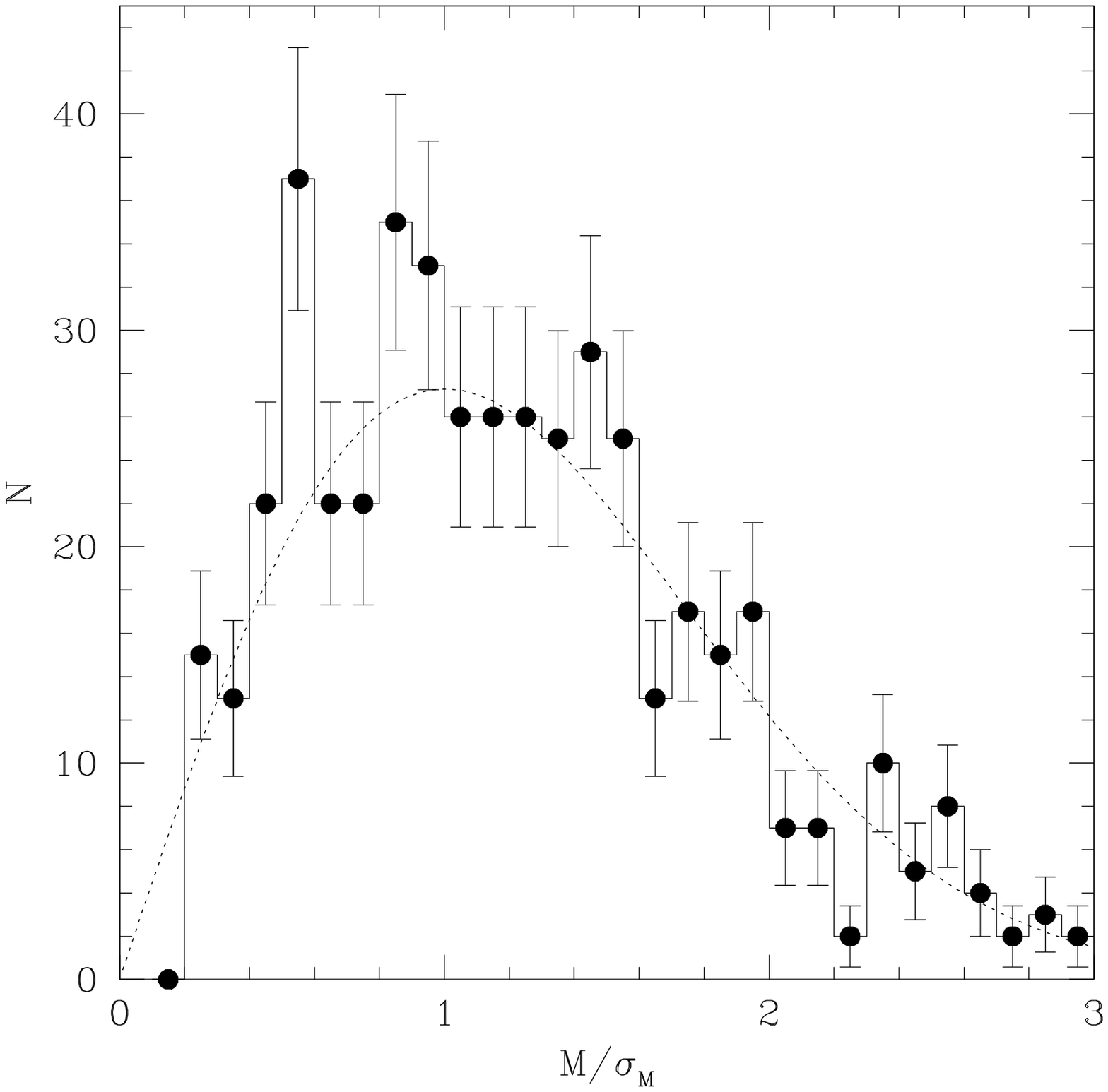}}
\caption{The solid histogram shows the distribution of planetary multiplicity
for one hundred realisations of our scenario, after 10~Myr (counting planets for which $a<1.1$AU). The dotted histogram is the best fit Poisson distribution for the number of planets in excess of two,
but it is not sufficiently narrowly peaked ($\chi^2 = 3.9$ per degree of freedom for bins $N_p$=3--8).
 The dashed histogram is the squared Poisson distribution
discussed in the text, which is a much better fit ($\chi^2=1.1$ per degree of freedom).}
\label{Multiplicity}
\caption{The solid histogram is the distribution of final planet masses, normalised by
the local value of $\sigma_M$, for planets
with $a<1.1$~AU. The dotted line is a Rayleigh distribution, in the scaled variable $M/\sigma_M$.}
\label{Mdis}
\end{figure}

Figure~\ref{Multiplicity} shows the distribution of true multiplicity for our
simulated ensemble, in which we count all planets that survive with semi-major axes $< 1.1$~AU, (this excludes a handful of surviving embryos on 
eccentric orbits, scattered out of the original disk), irrespective of mass. The median number of surviving planets, so defined, is four.

Statistical analyses of Kepler data with pre-defined functions often assume a functional form for the multiplicity that is based on the Poisson
distribution, (e.g. Fang \& Margot 2012). We find that this is not a good fit to our simulations because it is too broad (i.e. it contains too many systems with either fewer
or more planets than the median value). Furthermore, the minimum number of planets found in our simulations is three, so that our function should be adopted
to describe the number in excess of two. To that end, we find that a better empirical model of the true distribution is to square the Poisson function
for the number of planets in excess of two, i.e.
\begin{equation}
p(N) = \frac{\lambda^{2(N-2)} e^{-2 \lambda}}{\left((N-2)!\right)^2}
\end{equation}
where $\lambda=3.8$ yields the best fit. 


The reason for the minimum multiplicity is related to the requirements of conserving mass, energy and angular momentum while assembling an
extended distribution of mass into a handful of planets through mutual gravitational perturbations. We examine this issue further in \S~\ref{Assemble}.  

\subsection{Mass and Radius Distributions}



The masses of surviving planets exhibit a trend of increasing mass
with semi-major axis.
Thus, we fit the mass distribution as a Rayleigh distribution
with a dispersion that depends on semi-major axis, namely
\begin{equation}
f(m) = \frac{m}{\sigma_m^2} e^{-\frac{1}{2} \left( m/\sigma_m \right)^2}
\end{equation}
and
 \begin{equation}
\sigma_m = 7 M_{\oplus} (a/1 AU)^{0.6}.
\label{Mtrend}
\end{equation} This resulting fit is
shown in Figure~\ref{Mdis}.

\begin{figure}
\centerline{\includegraphics[width=0.52\textwidth]{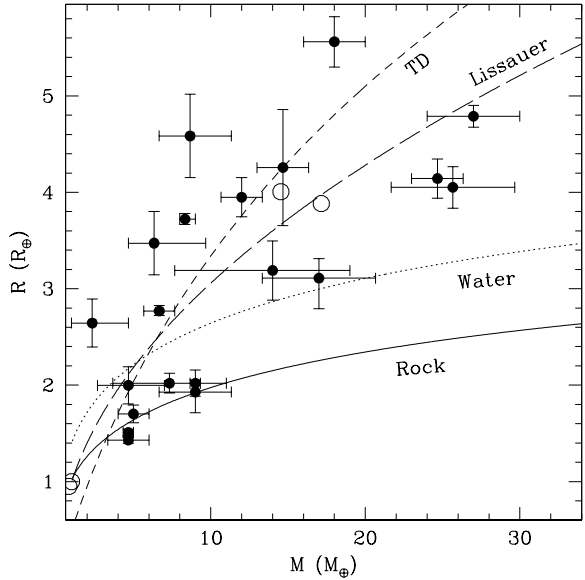},\includegraphics[width=0.52\textwidth]{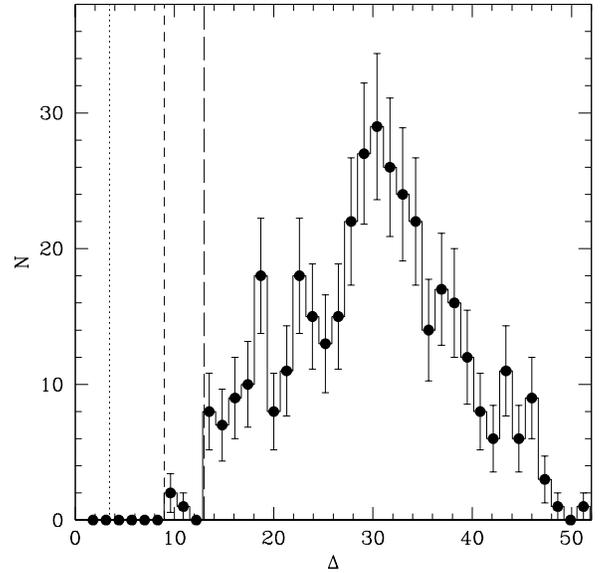}}
\caption{The solid points are masses and radii for extrasolar planets measured through transits and radial velocities, based
on the known planet sample as of Dec~2012. The
open points represent the solar system planets Venus, Earth, Uranus and Neptune.
The
solid and dotted curves are theoretical mass-radius relations from Seager et al. (2007), for Perovskite and Water respectively.
The long dashed line is the empirical mass-radius relationship from Lissauer et al. (2011b), and the short dashed line is the
relationship derived by Tremaine \& Dong (2012). Our radius estimates are based on the solid curve, with a potential multiplicative
factor $R'$ applied, to account for potential Hydrogen envelopes.}
\label{MR}
\caption{The solid histogram shows the distribution of $\Delta$ for the planet pairs with $a<1.1$AU.
The vertical dotted line is the stability limit for Jovian mass pairs derived by Gladman (1993). The vertical dashed line is for $\Delta = 9$
derived by Smith \& Lissauer (2009) for earth-mass systems. We see that our distribution cuts off completely at this value, although the more
accurate effective cutoff is $\Delta=13$, and the
bulk of the
pairs are more widely separated than this, peaking around $\Delta=30$, with few systems with $\Delta>50$.}
\label{Del}
\end{figure}

For comparison to the Kepler sample, we need to convert masses to radii.
In the simplest incarnation, our model assumes that the masses of the planets are
dominated by the rocky component. Therefore, we
 use the mass-radius relation of Seager et al. (2007), based on the
Perovskite equation of state, to characterise our planetary radii. 
 Figure~\ref{MR} shows the current sample of planets with both
mass and radius measurements. It is clear that many low mass planets have
radii in excess of that expected from our no-gas model, suggesting that the observed Kepler planet 
possess atmospheres of lower molecular weight material, either water or gas.

Previous authors have attempted to address this issue by interpolating
between rocky planets and more massive, gaseous planets. Lissauer et al. (2011b)
suggest a mass radius relation $ M = M_{\oplus} (R/R_{\oplus})^{2.06}$, based
on an interpolation between the solar system terrestrial planets and ice giants.
Tremaine \& Dong (2012), on the other hand, interpolate across the mass
range probed by transiting planets. Wu \& Lithwick (2012a) find an empirical
mass-radius relation $ M = 3 M_{\oplus} (R/R_{\oplus})$ by placing mass constraints
on planetary pairs using transit-timing variations.
 In each case, these relations amount
to an implicit, but not well characterized, assumption about the nature of the planetary
atmosphere. 

We adopt a radius based on the Seager et al. relationship,
but allow for a multiplicative radius enhancement factor $R'$, to account for
a potential gaseous atmosphere. This is a reasonable approximation for the radius
of a planet with a rocky core and hydrogen atmosphere of constant mass fraction,
at least in the regime (gas mass fractions $\sim 15\%$ or less) necessary for
 the known exoplanets with masses $< 10 M_{\oplus}$. A pure water planet
is also well described in this way, with an enhancement factor $R'=1.3$.

\subsection{Spacing of Planets}


The spacing of planets in multiple systems potentially contains vital information about
how the system was assembled. In HM11 we examined a variety of statistics commonly
used in assessing models of solar system formation, and found that the in situ assembly
model was able to reproduce the properties of the systems observed by Howard et al 
and Mayor et al. Table~\ref{OutTab} shows the same quantities for these simulations.

One of the most common measures for planetary spacing is $\Delta$, the orbital spacing
between two planets measured in terms of their average Hill radius
$$
\Delta = 2 \frac{a_2 - a_1}{a_2+a_1} \left( \frac{m_1 + m_2}{3 M_*} \right)^{-1/3},
$$
where $a_1,m_1$ and $a_2,m_2$ are the semi-major axes and masses of the inner and outer planets, and $M_*$ is the
mass of the host star.
Figure~\ref{Del} shows the ensemble distribution of $\Delta$ 
 in our simulations. 
 The distribution is broad, peaking at $\Delta \sim 30$. Various authors
have evaluated the lower limit on $\Delta$ imposed by dynamical stability, with $\Delta > 9$
being the expectation for earth-mass planets on initially circular orbits (Smith \& Lissauer 2009).
Our simulations do show a handful of systems just above this limit, but suggest that the more
significant edge to the distribution lies at $\Delta \sim 13$. At the other end, we find that
$\Delta > 50$ is rare, suggesting that transiting planet pairs with larger values likely
host additional, as yet undetected, planets.

\begin{figure}
\centerline{\includegraphics[width=0.52\textwidth]{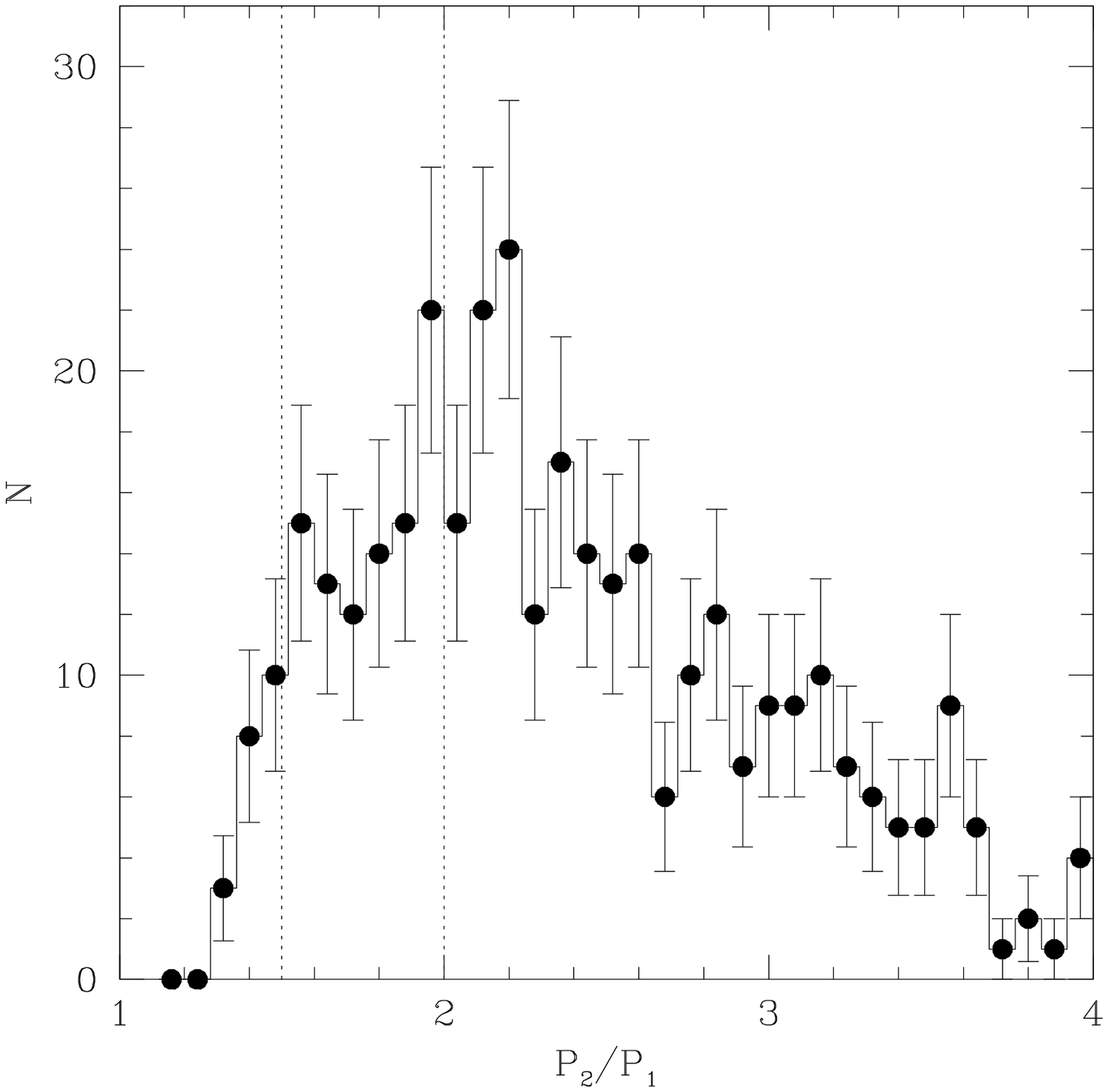},\includegraphics[width=0.52\textwidth]{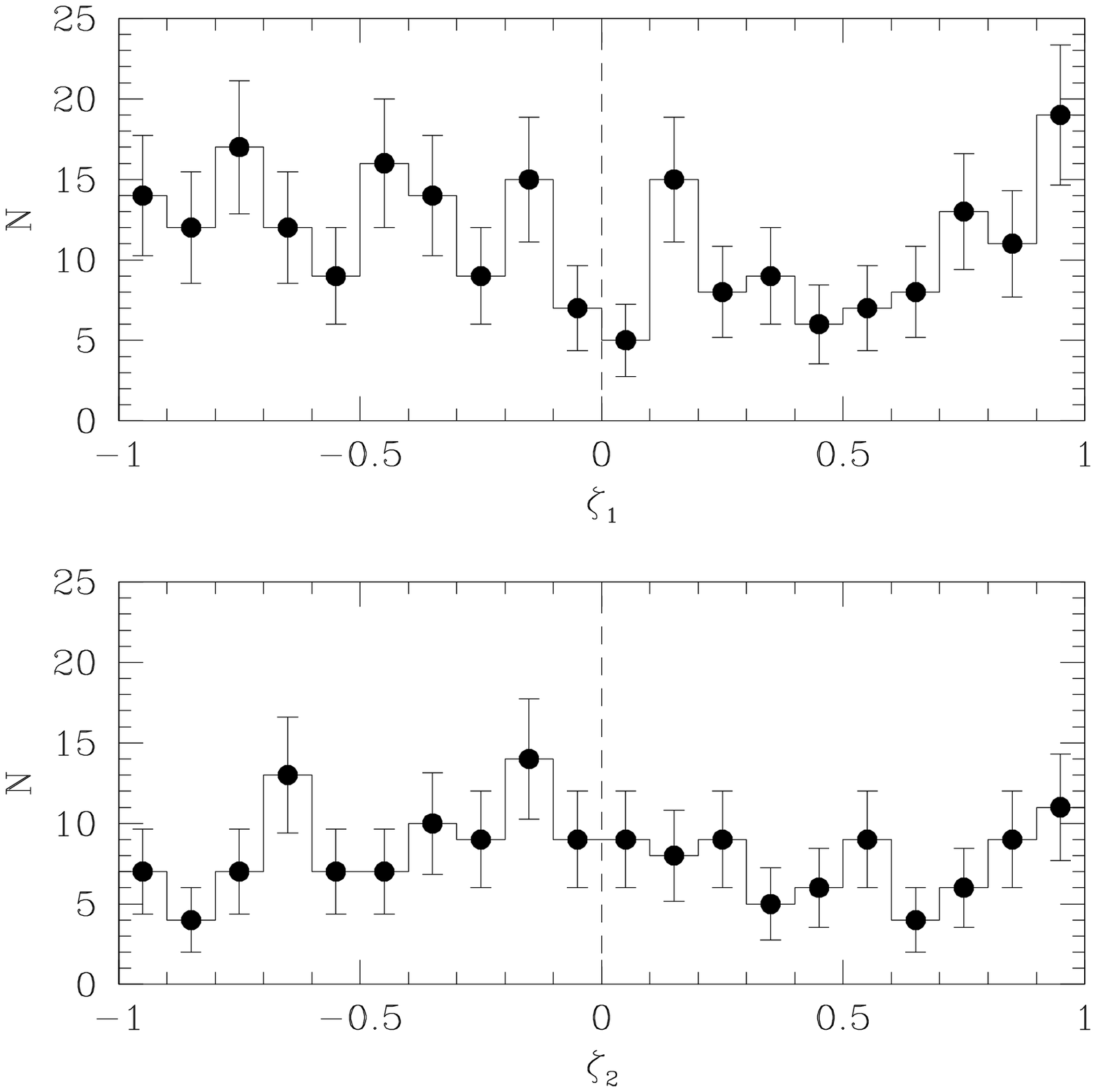}}
\caption{The observed distribution of period ratios between neighhouring pairs (in which
both lie within $a<1.1$AU). The dotted lines indicate the positions of the 2:1 and 3:2 orbital
commensurabilities.}
\label{Pbin}
\caption{The upper panel shows period ratios now binned in terms of $\zeta_1$, the proximity to
the nearest first order commensurability. Exact resonance is at $\zeta_1=0$ (shown by the vertical
dashed line). The surviving systems appear to avoid orbits too close to resonance. The lower panel shows
$\zeta_2$ (proximity to a second order resonance) for those systems for which $\left| \zeta_1 \right|>1$.
We see little indication of structure in this instance.}
\label{zbin}
\end{figure}

The statistical measures described in HM11, and also $\Delta$, are based on knowledge of the planetary
mass. For most Kepler candidates, however, we have only radius measurements. Thus, a more appropriate 
measure for the compactness of these systems is the distribution of period ratios in neighbouring
pairs. Lissauer et al. (2011b) and Fabrycky et al. (2012) have discussed this quantity and demonstrated
that the Kepler sample shows a broad distribution with additional structure near commensurabilities.
Figure~\ref{Pbin} shows the distribution from our simulations, peaking at ratios just about two. There
is some hint of structure near the 2:1 commensurability, but with neither the strength or the asymmetry
seen in the observational sample. 

Lissauer et al. (2011b) also define the $\zeta_i$ statistic, to measure the proximity to commensurabilities
of order $i$. Figure~\ref{zbin} shows the distributions of $\zeta_1$ and $\zeta_2$ for our simulations.
We see that there is some measure of avoidance of first order resonance in the simulations, and no
corresponding feature with respect to second order resonance. However, we again see a symmetric structure
rather than the asymmetry observed in Lissauer et al or Fabrycky et al. Nevertheless, the broader features
of the distribution are close enough that they may provide the initial basis on top of which additional
dynamical structure is created by dissipative effects such as tidal evolution, as proposed by Wu \& Lithwick (2012b) or Batygin \& Morbidelli (2013).

We can cast the same information in the direct form of a distribution of separations, which we call
$\delta$ (since it is essentially the un-normalised version of $\Delta$), shown in Figure~\ref{deltaprime}.
In this case, we also want to account for a slight trend of $\delta$ with semi-major axis, fitting the
distribution about this. The origin of the trend is the fact that annular spacing is dependant on the amount
 of mass in the disk, as discussed in \S~\ref{Assemble}. Thus, we normalise $\delta$ to $\delta_0$,
where
\begin{equation}
\delta_0=0.43-0.23 \log_{10} \left( \frac{a}{1 \rm AU} \right)
\end{equation}
and model the resulting distribution of $\epsilon=\delta/\delta_0$ as
\begin{equation}
f(\epsilon) \propto  e^{-\frac{1}{2} \left( (\epsilon - 0.82) / 0.07 \right)^2} +
0.6 e^{-\frac{1}{2} \left( (\epsilon - 1.28) / 0.12 \right)^2} +
0.48  e^{-\frac{1}{2} \left( (\epsilon - 0.53) / 0.1 \right)^2} +
0.56  e^{-\frac{1}{2} \left( (\epsilon - 0.98) / 0.05 \right)^2}.
\end{equation}
This is the function we will use to determine the spacing of our planetary systems. However, since the
mass and separations are chosen seperately, we will include an additional step of removing systems from
the Monte Carlo model that violate the requirements $13 < \Delta <50$.

\subsubsection{An approximate theory of planet spacings}
\label{Assemble}

It is of interest to understand how the spacing and multiplicity of these systems
is determined. The underlying process is
 the gravitational assembly of
an extended mass distribution, regulated by both the requirements of global mass, energy
and angular momentum conservation, as well as the requirement that the assembled bodies
are sufficiently massive to perturb and eventually accrete any material nearby.

The global conservation requirements are simple. Although gravitational perturbations can
potentially eject material from a planetary system, the systems we simulate are sufficiently
deep in their stellar potential wells that the amount of material lost via this mechanism is negligible,
and similarly for accretion onto the star. Thus, for a disk with surface density $\Sigma = \Sigma_0 (r/r_0)^{-\alpha}$,
spread from $r_{in}$ to $r_{out}$, the total mass, energy and angular momentum budgets are

\begin{eqnarray}
M_{tot} & = & 2 \pi \Sigma_0 \frac{ r_0^{2} \left( (r_{out}/r_0)^{2-\alpha} - (r_{in}/r_0)^{2-\alpha}\right)}
{2 - \alpha} \\
E_{tot} & = &  - \pi G M_* \Sigma_0 \frac{ r_0 \left( (r_{out}/r_0)^{1-\alpha} - (r_{in}/r_0)^{1-\alpha}\right)}
{1 - \alpha} \\
L_{tot} & = & 2 \pi \left( G M_* \right)^{1/2} \Sigma_0 \frac{ r_0^{5/2} \left( (r_{out}/r_0)^{5/2-\alpha} - (r_{in}/r_0)^{5/2-\alpha}\right)}
{5/2 - \alpha} \\
\end{eqnarray}

If this were to be incorporated into a single final planet, of mass $M_{tot}$, then energy conservation dictates that the final
semi-major axis is 
\begin{equation}
 a = r_0 \frac{1-\alpha}{2-\alpha} \frac{ \left( (r_{out}/r_0)^{2-\alpha} - (r_{in}/r_0)^{2-\alpha}\right)}
 { \left( (r_{out}/r_0)^{1-\alpha} - (r_{in}/r_0)^{1-\alpha}\right)}
\end{equation}
and angular momentum conservation requires that the eccentricity be given by
\begin{equation}
1 - e^2 = \frac{ (2 - \alpha)^3}{(\alpha-1) (5/2-\alpha)^2}
\frac{(1 - y^{1-\alpha}) (y^{5/2-\alpha}-1)^2}{(y^{2-\alpha}-1)^3}
\end{equation}
where $ y = r_{out}/r_{in}$.
It proves impossible to satisfy these two conditions unless $y=1$ (i.e. unless we specify an infinitely narrow ring, i.e. a single planet),
for $\alpha>0$. This is because each annulus of the disk is assumed to be on a circular orbit and therefore carries the maximum angular momentum
per unit energy. The energy integral is a harmonic average, which shifts the final location for the putative single planet inside the median radius
of the disk, resulting in an excess of angular momentum per unit energy if one assumes all the mass is accumulated into a single planet. Thus, we require at least two planets to 
conserve all of the mass, energy and angular momentum simultaneously in the assembly process.

In addition, this assembly requires that the final planets be massive enough to gravitationally perturb and accrete all the material in the
annulus into one or other of the final planets. If the original disk is not sufficiently massive, the final state may naturally consist of 
additional planets. We can estimate this, at least for the size of disk we simulate, by noting that our simulations suggest that $\Delta \sim 50$
is an upper limit to the degree of separation of final planets. If an annulus produces two planets at the inner and outer edges of the annulus, and
their mutual $\Delta > 50$, then it is likely that a third, intermediate planet forms from material that is too far from either of the other two planets
to be perturbed onto a crossing orbit. We can express $\Delta$ as
\begin{equation}
\Delta = 73.7 \frac{y-1}{y+1}  \left( \frac{M_{tot}}{20 M_{\oplus}} \frac{y^{1/2}-1}{3.472} \left(\frac{r}{0.05 \rm AU} \right)^{1/2} \right)^{-1/3}
\label{Deltaeq}
\end{equation}
where $r$ and $y$ are now running variables representing the inner edge and width of a particular annulus corresponding to the `feeding zone' of
a particular final planet. The number 3.472 represents the quantity $y^{1/2}-1$ for the entire original disk from 0.05~AU to 1~AU, and $M_{tot}$
represents the mass of the entire disk, normalised by a host star of mass $1 M_{\odot}$. We have also assumed $\alpha=3/2$ for this case.

Thus, as an example, assuming that one planet forms at the inner edge of the disk at 0.05~AU, the requirement that $\Delta < 50$ suggests $y \leq 2$ under
these conditions, so that the second planet lies no further than $r=0.1$~AU. Using this location as the inner edge of the next annulus suggests $y \sim 2.4$
and a third planet at 0.24~AU. Similarly, this indicates $y \sim 3.5$ and a fourth planet at 0.85~AU, which is close to the outer edge of
our disk. Thus, we expect at least four planets
to form from our original disk and the simulations do indeed produce mostly four and five planet systems. If we repeat this exercise with
$\Delta \sim 30$ (closer to the median of the distribution), the values of y range from 1.5--2, and one can fit 6 planets between 0.05~AU and
1~AU, and 7 if the outer edge is 1.2~AU. Thus, stability considerations suggest that most systems should have between 4 and 7 planets, as is
indeed observed. Note that the range of y encompassed by these arguments does allow for the presence of 2:1 period ratios. Even ratios in the
3:2 range are dynamically allowed, although they require $\Delta$ values between the median and the stability limit.

\begin{figure}
\centerline{\includegraphics[width=0.52\textwidth]{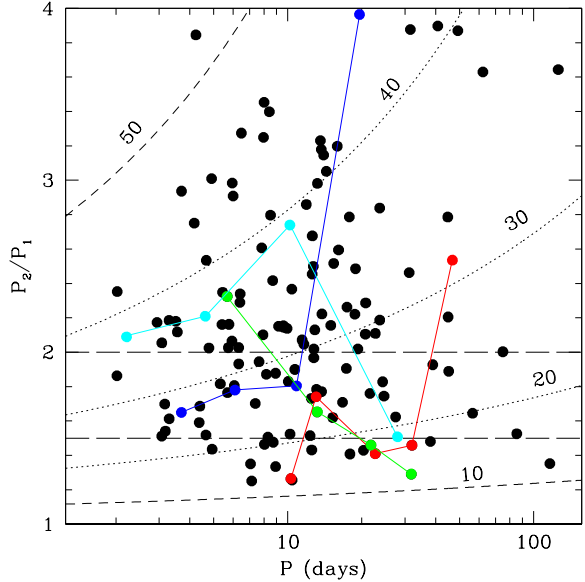},\includegraphics[width=0.52\textwidth]{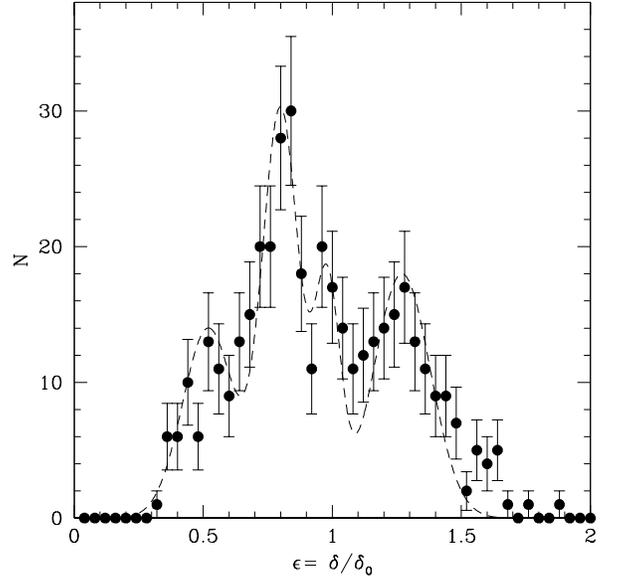}}
\caption{The solid points represent Kepler planetary candidate pairs, plotted as period ratio versus the orbital period of the 
inner member of the pair. Coloured points and chains represent successive planets in the same system. The systems shown are all ones with
one or more confirmed or validated planet -- Kepler-11 (red); Kepler-20 (blue); Kepler-33 (green) and Kepler-45 (cyan). The curved dotted
lines indicate curves of constant $\Delta$, calculated assuming the relation in equation~(\ref{Deltaeq}). The horizontal dashed lines indicate
the period ratios associated with the 2:1 and 3:2 commensurabilities.}
\label{KepDel}
\caption{The distribution of true (i.e. not weighted by mass) spacings of neighbouring planets
shows several broad peaks. The distribution shown here has had a mean trend divided out. The dashed line
shows the fitting function we use to characterise this distribution.}
\label{deltaprime}
\end{figure}

One can convert the quantity y into a period ratio for each planet pair. Figure~\ref{KepDel} shows the distribution of period ratios as a 
function of inner planet period for all planets in multiple systems that fall within our sample cuts in \S~\ref{Obs}. Also shown are lines
of constant $\Delta$ using equation~(\ref{Deltaeq}). This is another way of representing the distribution shown in Figure~\ref{Del}, and we find 
that indeed the bulk of the planets cluster in the range $20 < \Delta <40$ and that the limits at 10 and 50 are good delimiters of the edge
of the distribution. Note that the observed data are tabulated only using periods, so that the uncertainties in the mass-radius relation are
not relevant.
 Figure~\ref{KepDel} also shows coloured chains for a handful of confirmed, high multiplicity Kepler systems. We see
that the outer pair of the Kepler-20 system (blue) has a period ratio that puts it in the high $\Delta$ wing of the distribution, suggesting
that an additional planet may be waiting to be found between Kepler-20f and Kepler-20d. Although the period spacing between Kepler-11f and Kepler-11g
deviates somewhat from that of the inner planets in the system, it is still well within the expected distribution, so that we do not necessarily
have a strong motivation to expect an additional planet in that gap.

\subsection{Eccentricities and Inclinations}

\begin{figure}
\centerline{\includegraphics[width=0.52\textwidth]{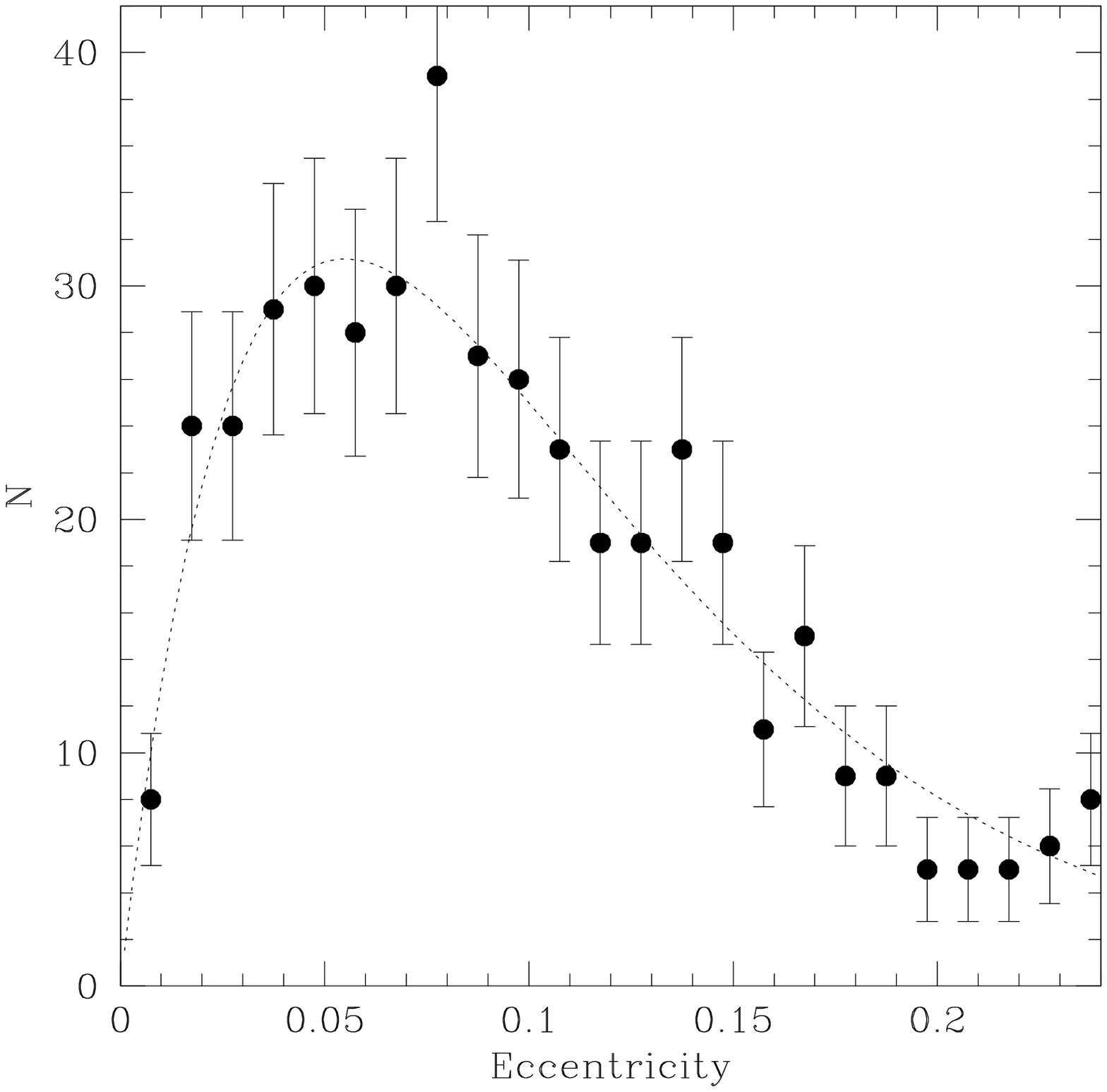},\includegraphics[width=0.52\textwidth]{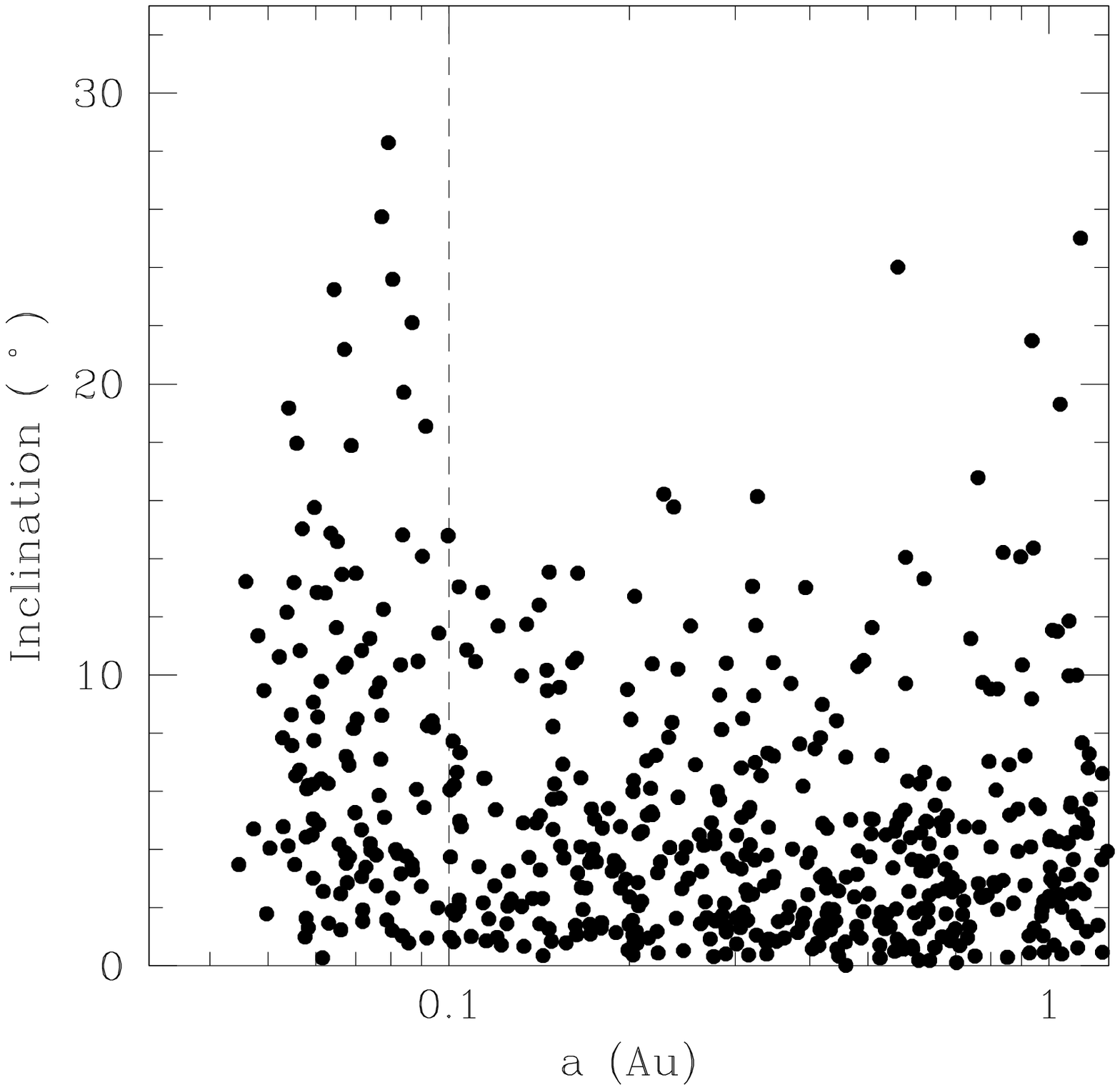}}
\caption{The solid points are eccentricities for all surviving planets interior to 1.1~AU. The dotted line is our exponential Rayleigh
distribution, with $\sigma_e=0.055$, giving a mean eccentricity of 0.11.}
\label{Ebin}
\caption{We show here the inclinations, relative to the original orbital plane, of all surviving planets in our 100 realisations
of the model. We see that the range of inclinations at semi-major axis $<0.1$AU is somewhat larger than those for $0.1{\rm AU} < a < 1$AU, so
we will fit to those two distributions seperately.}
\label{ia}
\end{figure}

Our initial conditions place all the planetary embryos on circular orbits, but
the process of assembly generates eccentricities by both secular interactions and
direct scattering and collision (e.g. Chambers \& Wetherill 1998). The resulting distribution of planetary eccentricities
is given by Figure~\ref{Ebin}. We see that the tail of large eccentricities is
somewhat larger than expected for a Rayleigh distribution (the normal assumption for
the functional form). A better fit is to replace the Gaussian function in the fitting
formula with a simple exponential, so that
\begin{equation}
 f(e) = \frac{e}{\sigma^2_e} e^{-e/\sigma_e} \left( 1 - \left( 1 + \frac{1}{\sigma_e} \right) e^{1/\sigma_e} \right)^{-1}
\end{equation}
with $\sigma_e = 0.055$. This results in a mean eccentricity of 0.11, broadly consistent with the
analysis of Moorhead et al. (2011) based on the distribution of Kepler transit durations.

The underlying eccentricity distribution will also be affected by tidal circularisation for
those systems with sufficiently small semi-major axis. In the absence of a more detailed
model for the dissipation in these low mass planets, we can adopt the traditional parameterisation
and estimate the circularisation semi-major axis as
\begin{equation}
a_{circ} = 0.15 {\rm AU} \left( \frac{T_*}{5 Gyr} \right)^{2/13} \left( \frac{R_p}{2 R_{\oplus}} \right)^{10/13}
\left( \frac{M}{5 M_{\oplus}} \right)^{-2/13} \left( \frac{Q}{100} \right)^{-2/13}
\end{equation}
where $T_*$ is the age of the host strs, $R_p$ and $M_p$ are the masses of the planets and
$Q$ is the quality factor of the dissipation in the planet. Thus, values of $Q \sim 10$--1000 will
place $a_{circ} = 0.1$--0.2~AU.

\begin{figure}
\centerline{\includegraphics[width=0.52\textwidth]{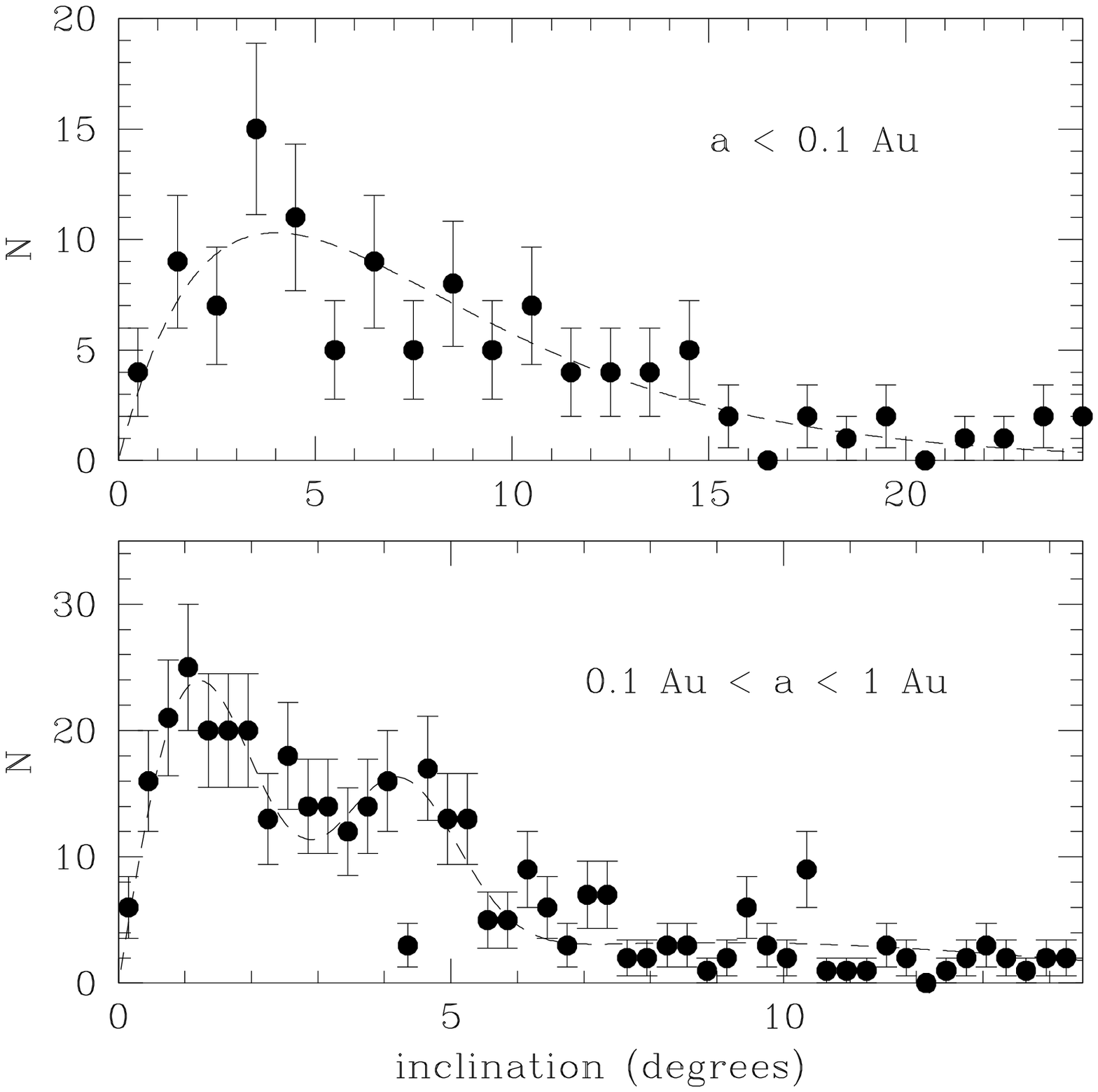},\includegraphics[width=0.52\textwidth]{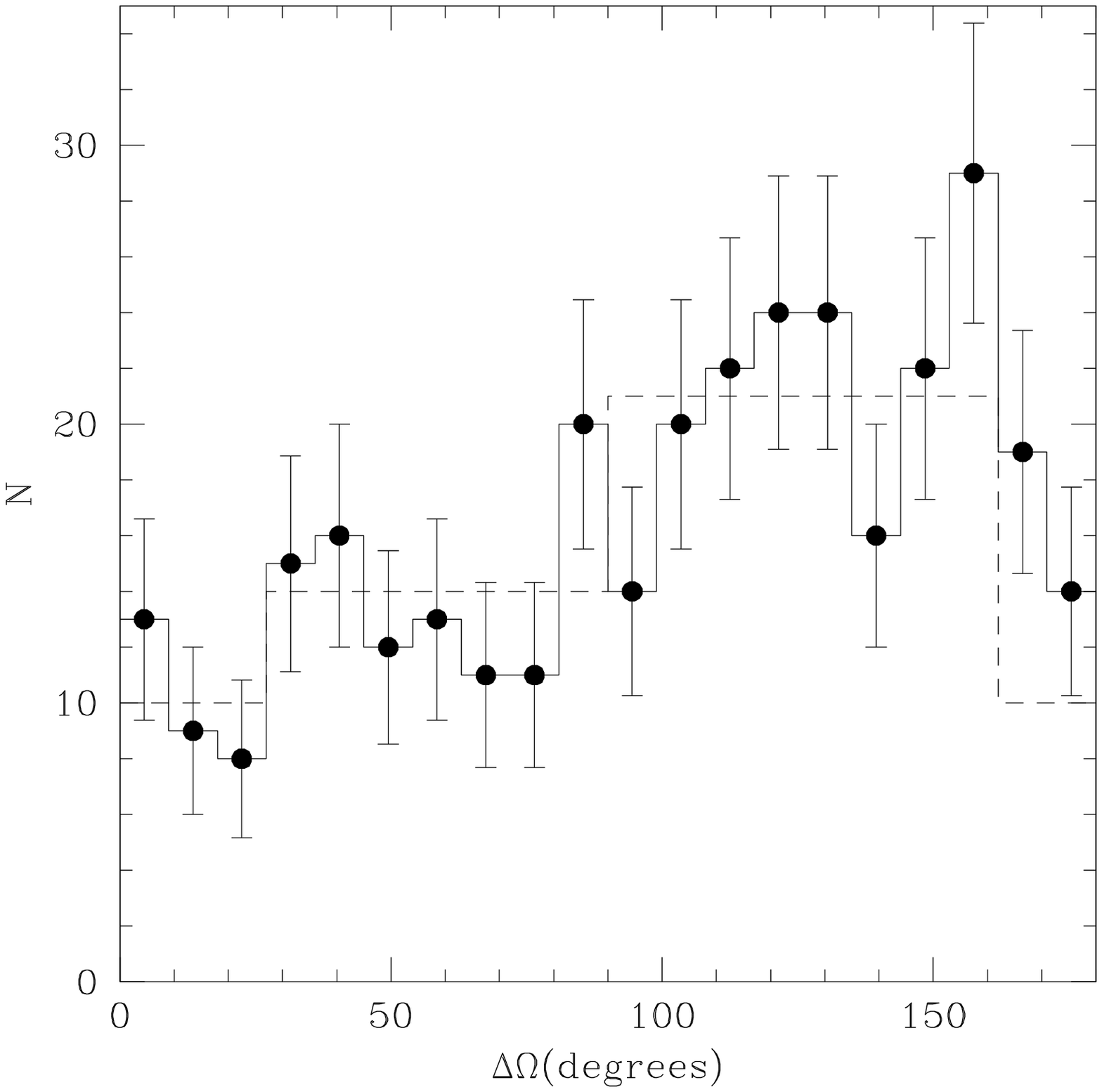}}
\caption{The points in the upper panel show the distribution of inclinations (relative to the original orbital plane) of all planets inside 0.1~AU.
The points in the lower panel show the same quantity for $0.1 {\rm AU} < a < 1.1 {\rm AU}$.
The dashed lines show the models described in the text.}
\label{ibin}
\caption{The binned data show the difference in the longitude of the ascending node for neighbouring pairs of planets. There is
a slight avoidance of orbits that are too close to aligned and a slight preference for $\Delta \Omega \sim 90^{\circ}$--$160^{\circ}$. We model this using the simple distribution given by the dashed line.}
\label{Delom}
\end{figure}

The inclinations of the planetary orbits are of  particular interest in the case of Kepler comparisons,
since they will determine what fraction of a system's planets actually transit the host
star from a given observational direction.
The ensemble of inclinations (relative to the original orbital plane) is shown
in Figure~\ref{ia}. We see that there is a trend for larger inclinations at small
semimajor axis, so we model this as two separate inclination distributions, shown
in Figure~\ref{ibin}. For semi-major
axes $a < 0.1$AU, we model the distribution as
\begin{equation}
f_1(i) \propto i e^{-i/4^{\circ}}
\end{equation}
while, for $a>0.1$AU, we model it as
\begin{equation}
f_2(i) \propto \frac{i}{1^{\circ}} e^{-\frac{1}{2} \left( i / 0.9^{\circ} \right)^2}
 + 0.35 e^{-\frac{1}{2} \left[ \frac{i - 3.9^{\circ}}{0.9^{\circ}} \right]^2}
 + 0.008 \frac{i}{1^{\circ}} e^{-\frac{1}{2} \left( i / 10.3^{\circ} \right)^3}.
\end{equation}
We see that this latter distribution contains a component corresponding to a Rayleigh
distribution with a dispersion similar to those determined by previous analyses of
the Kepler data (e.g. Lissauer et al. 2011b, Fabrycky et al. 2012, Tremaine \& Dong 2012, Fang \& Margot 2012).
However, it also contains a higher inclination component which may bias the inferences
based on an assumed distribution of the Rayleigh form. Similarly, the inclination distribution
for the closest planets has a larger tail to high inclinations than that of a Rayleigh distribution.

The inclination distribution describes the orientation of the orbital plane relative
to that of the original disk. In order to fully describe the orientation relative to 
a given observer, we also need to specify the angle of the line of nodes. Figure~\ref{Delom}
shows the offsets in this angle, $\Omega$ for neighbouring pairs of planets, calculated in
the original orbital plane of the disk. We see that
the orientations of neighbouring orbits are not completely uncorrelated, and that there is a
tendency to avoid orbits that are exactly aligned with each other. We model this with the
simple piecewise distribution shown in Figure~\ref{Delom}.

\subsection{Monte Carlo Model}
\label{MC}

Kepler has detected thousands of planets in several hundred systems, which represent a sampling
of a potentially much larger parameter space. Simulating such a data set directly is unrealistic,
but we can use the parameter distributions of the previous sections to construct a Monte Carlo
model for the generation of planetary systems. This will produce model populations with the
same properties as the simulations, except for the exclusion of any parameter correlations 
(such as the mass-period trend and nodal distribution) not
explicitly identified. 

We draw planetary separations from the distribution function $f(\epsilon)$ and planetary
masses from $f(m)$ (out to distances of 1~AU). We then evaluate the value of $\Delta$ for each pair and reject
any systems containing a pair that violates $\Delta<13$ or $\Delta>50$ (this is a small minority of cases
because the distribution of $f(\Delta)$ is closely related to $f(\epsilon)$ and $f(m)$).
We also draw distributions of inclination and eccentricity from our functions $f(e)$ and $f_1(i)$
or $f_2(i)$. We convert masses to radii using the Seager et al. equation of state and
the value for $R'$.

The sample of true underlying planetary systems is then `observed' by randomly choosing a particular
plane of observation and calculating the inclinations of all planetary orbits relative to
that plane. We also account for the distribution of the planetary lines of nodes with respect
to a particular direction chosen for the observer. All planets whose impact parameters are $< 1 R_{\odot}$
are then counted as transiting. In order to avoid confusion, we will refer to these as `tranets', in the
parlance of Tremaine \& Dong (2012).

In order to match with the observations, we must also reject tranets whose radii are too small
to be detectable. In the interests of clarity, we will restrict our comparison to the parameter
range where more detailed studies (Howard et al. 2012; Youdin 2011) suggest that the completeness
is high.
 These prior studies suggest that the Kepler database
is relatively complete for $R > 1.5 R_{\oplus}$ and $P < 200$~days, and so we adopt this as our
conservative observational limit criteria.

\section{Comparison to Observation}
\label{Obs}

In order to compare our Monte Carlo model to the observations, we adopt the following criteria (following 
 Howard et al. 2012 and Youdin 2011) to define a complete
subsample of the publicly released Kepler planet candidate catalog as of Jan 2013 (although
we also compare to the sample 
announced in Batalha et al. (2012) in case there have been any systematic changes in sample selection
with time).
 Since our simulations are for systems
around a $1 M_{\odot}$ star, we
 adopt $4100 K < T_{eff} < 6100$K and $\log g>4$ in order to approximately match the stellar
hosts to the model. We also
 require Kepler magnitudes $<15$ to ensure that the completeness is not reduced because of
lower photon statistics and signal-to-noise.
 We also then require $P<200$~days and $1.5 R_{\oplus}< R < 10 R_{\oplus}$.
This leaves us with 463 single tranet systems, 79 two tranet systems, 26 three tranet systems, 7 four tranet systems,
 1 five tranet system and 1 six tranet system (For the Batalha et al. (2012) catalog, the numbers are 463:85:25:2:2:1).

\subsection{Mutual Inclinations}

\begin{figure}
\centerline{\includegraphics[width=0.52\textwidth]{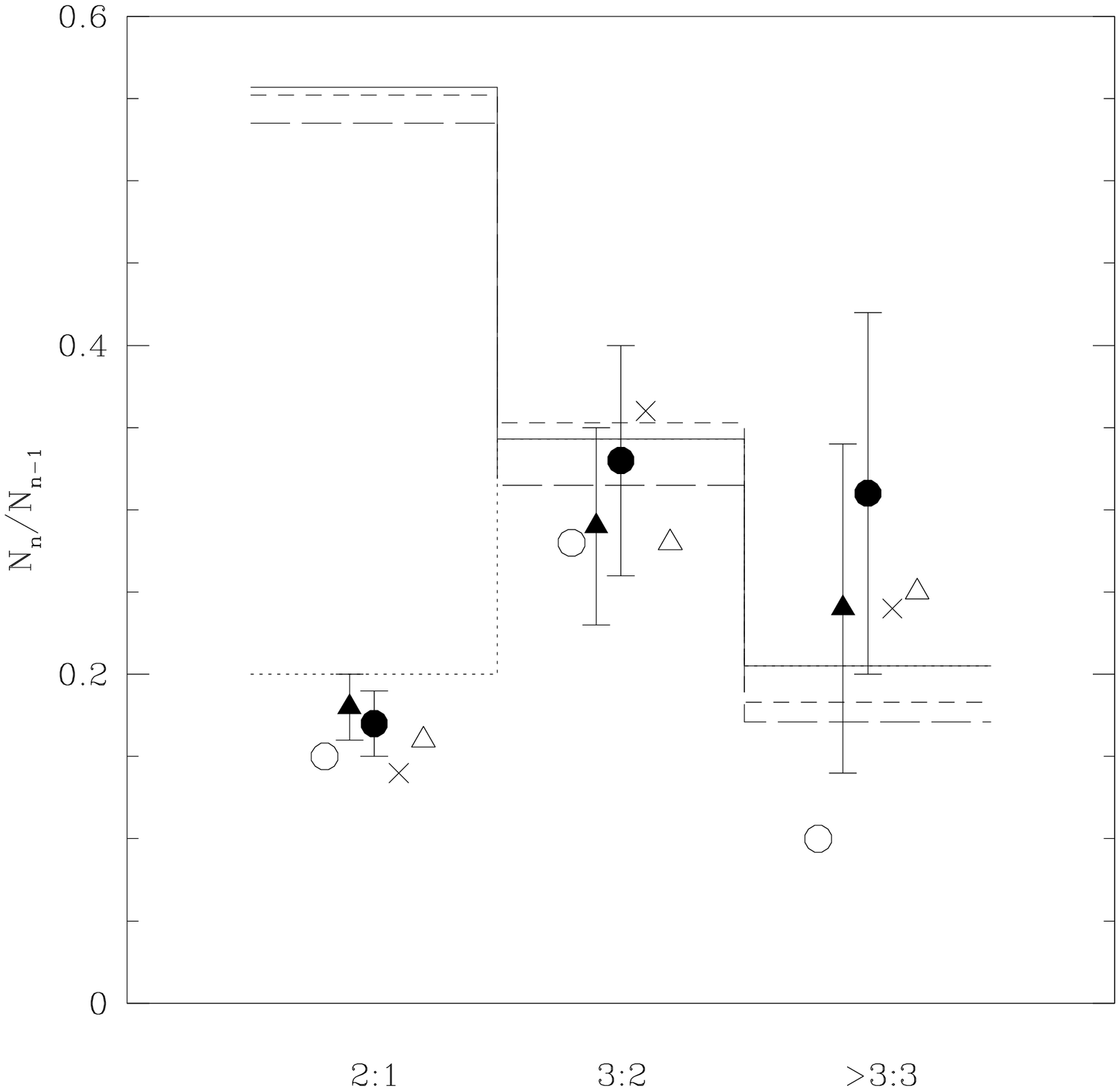},\includegraphics[width=0.52\textwidth]{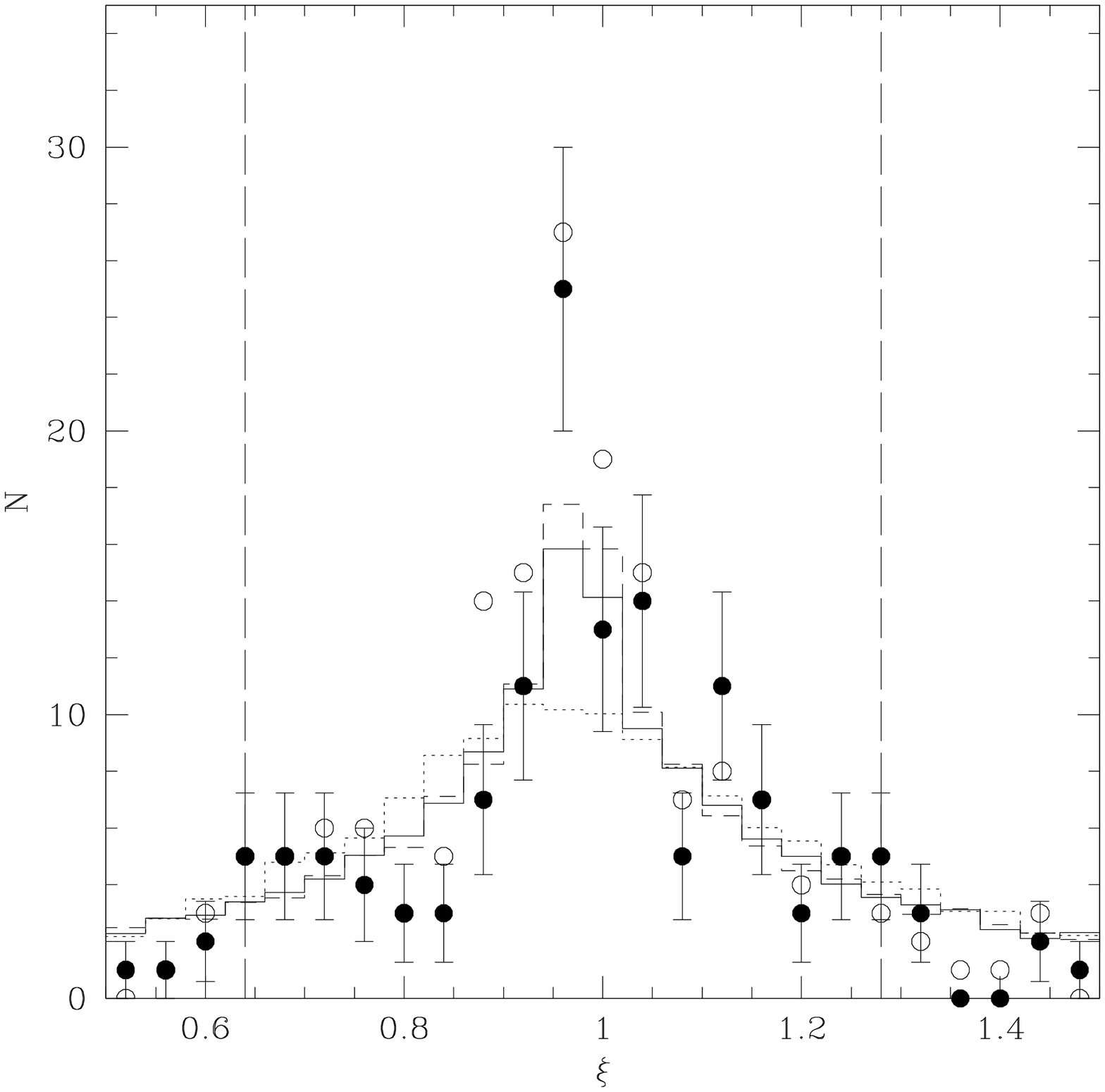}}
\caption{The solid points, with error bars, represent the multiplicity ratios from the Kepler sample as defined in the text. The circles
are drawn from the Jan 2013 Kepler catalogue, while the triangles are drawn from the previous incarnation (Batalha et al. 2012).
Other symbols show samples defined by other authors from Kepler data.
 The open
circles are those from the study of Lissauer et al. (2012) based on the first major sample of multiple tranets. The crosses are for the sample determined by Tremaine \& Dong (2012) and the open triangles are those of Fang \& Margot (2012) (Figuera et al. 2012 are also consistent, although not plotted because they didn't account for five or six tranet systems). The consistency of all such samples indicates that this comparison is relatively robust with regards to sample selection.
 The relative horizontal offsets are arbitrary and for the purposes of clarity. The solid histogram represents the expected ratios from our Monte Carlo model with a radius enhancement
 of $R'=1.2$. The dashed histograms correspond to the same model but with $R'=1.0$ (long dashes) and $R'=1.4$ (short dashes). We see that the model robustly reproduces the multiple tranet ratios and is not heavily
sensitive to $R'$. 
In all cases the ratio of singles to doubles is dramatically undercounted. The dotted histogram shows the equivalent of the solid 
histogram except that we have assumed that another process produces single tranet systems only, at a rate equivalent to that which produces the multi tranet systems.}
\label{MultiStats}
\caption{The solid points are the observed distribution of $\xi$ for the sample defined in the text from the Jan 2013 catalog. The
open points are defined in the same way but using the earlier sample from Batalha et al. (2012). The dotted histogram represents the distribution
from our model, with $R'=1.2$ and using the full eccentricity and inclination distributions. The dashed
histogram represents the model result if we keep the same inclination distribution but set all eccentricities to zero. The solid histogram
represents an intermediate model in which orbits are circularised only 
 for $a<0.25$AU.
 Note that all points outside the vertical dashed
lines are excluded from the fit because of low number counts.}
\label{xiplot}
\end{figure}

One of the simplest measures of comparison between data and observation is the ratios of systems of different multiplicity.
Figure~\ref{MultiStats} shows the comparison of observations with the statistics of our Monte Carlo model for different values
of the radius enhancement $R'$, as well as equivalent measures from several other studies in the literature. We see that
the models can reproduce the ratios of triples to doubles and of triples to systems with four or more tranets. We also see
that this consistency is robust to uncertainties in $R'$, as the
predicted ratios don't change much over a range of $R'$ consistent with the known planetary mass-radius relation. 
There is also a consistency of the observed ratios despite different sample selections by different authors
(Lissauer et al. 2011b; Tremaine \& Dong 2012; Fang \& Margot 2012), suggesting that the comparison is also robust
with respect to the particular observational cuts.

However, our models fail to reproduce the observed ratio of multiples to single tranets, in the sense that the Monte Carlo
model severely underpredicts the number of systems which show a single transiting system. Lissauer et al. (2011b) noted the
potential excess of singles and argued that simple geometric considerations suggest that the ratio of singles to doubles should
be higher unless there is an unusual correlation or inclination dispersion. To that end, we paid special attention to modelling
the high inclination portion of the distributions in Figure~\ref{ibin} and to potential correlations in the lines of nodes, but
neither of these appears to have materially affected the results. Although other studies such as Tremaine \& Dong (2012) and
Fang \& Margot (2012) claim consistency in their modelling of all the data, this is because their chosen statistical distributions 
allow for a substantial population of single transiting systems, something not allowed by our more physically derived model.

We can, of course, simply postulate a second unknown process at work that produces only single tranet systems, which can then be combined
with our model. If we postulate that this second process operates at the same rate as our model, we predict a ratio of singles
to doubles $\sim 0.2$, which matches the observations. 

A second, more accurate measure of the mutual inclination distribution is the velocity-normalised transit duration ratio 
\begin{equation}
\xi\equiv {T_{\rm dur,1}/P_1^{1/3}\over T_{\rm dur,}/P_2^{1/3}},
\end{equation}
where $T_{\rm dur}$ is the transit duration and $P$ is the orbital period (Fabrycky et al. 2012). This is an empirical measure of the ratio of the impact parameters for 
two neighbouring tranets,
which should scale as the semi-major axis ratio for perfectly aligned systems. Figure~\ref{xiplot} shows the distribution of $\xi$
that results from our Monte Carlo model, compared to that of the observed sample as defined above. The dotted histogram is for
$R'=1.2$, using the eccentricities and inclinations drawn from the simulated systems. We see that the model distribution captures
the wings correctly, but underpredicts the peak near $\xi \sim 1$. This behaviour is the same for all values of $R'$. The dashed
histogram shows the model distribution if we use the same $R'=1.2$ and full inclination distribution, but circularise all the orbits.
We see that this model captures the observations
very well, with a $\chi^2$ per degree of freedom of 1.15. Indeed, reasonable agreement can be obtained by only circularising orbits
within a certain upper semi-major cutoff. The solid histogram shows the $\xi$ distribution for the case in which this limit
is 0.20~AU, which defines the $2 \sigma$ level with respect to the fit obtained with full circularisation.
 We note that the improvement in the fit is solely due to the reduction in the eccentricity -- we do not assume any change
in the mutual inclinations as a result of tidal effects.

\subsection{Period Distributions}

\begin{figure}
\centerline{\includegraphics[width=0.52\textwidth]{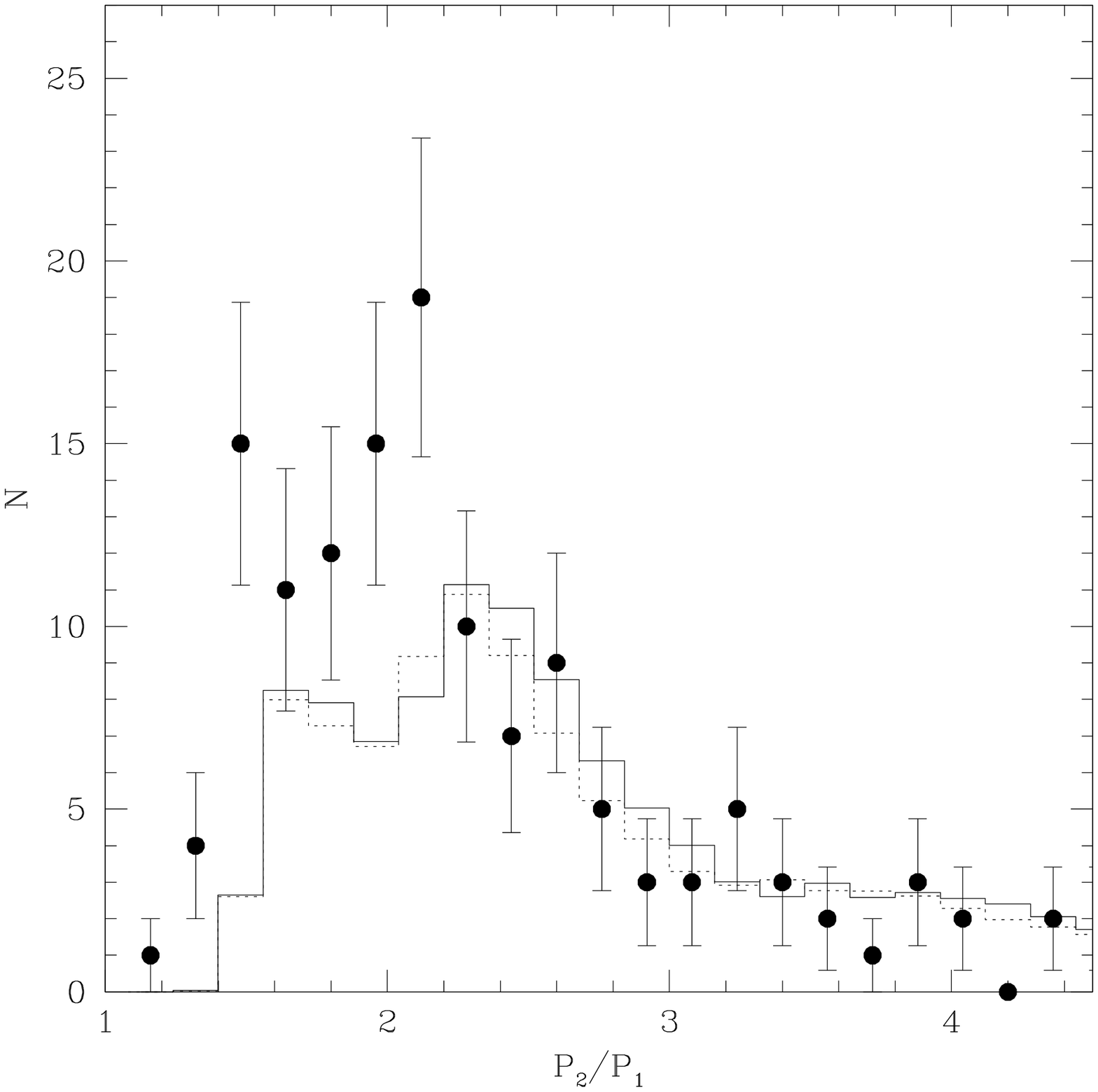},\includegraphics[width=0.52\textwidth]{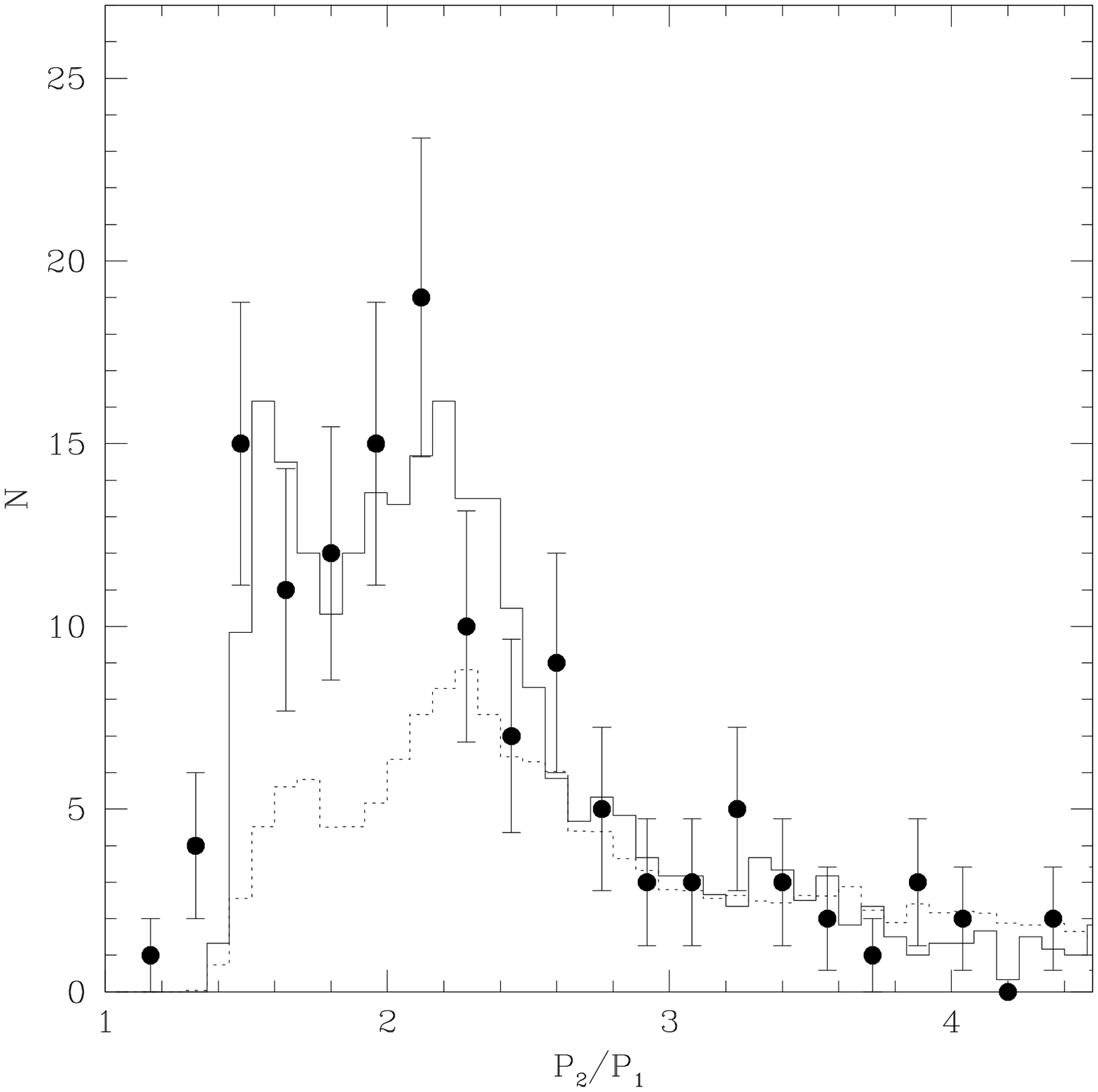}}
\caption{Solid points represent the period ratio distribution of our observational sample. The solid histogram
is the corresponding distribution from the Monte Carlo model with $R'=1.2$. The dotted histogram is for the same parameters
but drawn from the model with an adjusted inner edge (\S~\ref{Edge}).}
\label{Prat}
\caption{The solid points are the same as for Figure~\ref{Prat}, but the solid histogram now represents only the
period ratios in systems where four or more tranets are detected. The dotted histogram represents the period ratios of systems in
which two tranets are detected. The distribution of high multiplicity systems better shows the two peaked structure evident in
the observations, albeit shifted slightly to longer orbital periods.}
\label{Prat2}
\end{figure}

The comparison of the model period ratio distribution to observations, shown in Figure~\ref{Prat}, is not quite as good as the various
measures that depend on inclinations. Some of the structure is reproduced, but the simulations underpredict the number of close pairs
relative to the number of more widely separated pairs. Furthermore, although there is a double peaked structure to the distribution
between period ratios of 1--3, the peaks are shifted to larger values than observed. Indeed, the entire shape of the distribution appears
shifted by $\sim 5$\% to larger values.
 This suggests that some level of dissipation is required beyond the simple assembly model assumed
here. It has been suggested that such enhancements near commensurabilities, and the slight shift in the observed
values, could be the result of tidal evolution (Wu \& Lithwick 2012b;
Batygin \& Morbidelli 2013). We have not implemented this effect, although we have included the semi-major axis shift associated with
circularisation for values less than 0.20~AU.

Some of this mismatch may also be due to poorly quantified selection effects. Steffen (2013) has noted that the period ratio distribution is
subtly different between high multiplicity (four or more tranet candidates) and low multiplicity (two tranet candidates) systems. In Figure~\ref{Prat2}
we show the corresponding comparison between the same data as in Figure~\ref{Prat} but now with two distributions drawn from the model in the same
spirit as Steffen (2013). We see that the high multiplicity model systems actually do a much better job of reproducing the data than the lower multiplicity
systems. The mismatch in Figure~\ref{Prat} is clearly the result of the stronger weighting towards the two tranet systems as in Figure~\ref{MultiStats}, but
the structure here suggests that selection effects that favour more high multiplicity systems could potentially improve the comparison as much as
moderate radial migration. We note also that the distribution from the two-tranet systems is indeed shifted slightly compared to the higher multiplicity
systems, as suggested by the comparison by Steffen.

We can also compare our simulations to the distribution of absolute periods of detected tranets. This is shown in Figure~\ref{Pdis}. We
see that the shape of the distribution is well reproduced for $P>$10~days, but that the simple model overproduces short period tranets
by a factor of three near the peak of the distribution. This is likely a consequence of our rather naive application of a single model realisation
to the full population, in the sense that our planetary embryo disks all have an inner edge at 0.05~AU and therefore almost always produce a
planet with an orbital period $<10$~days. If the true underlying model for the planetary initial conditions contains some variation in the
location of the inner edge, then this peak will be reduced.

\subsection{Model Embellishments}
\label{Edge}

We have kept our default model simple in order to highlight the robustness of the match to the data, but we can also demonstrate
that a more comprehensive match to the data is achievable with minor modifications to the model.

\begin{figure}
\centerline{\includegraphics[width=0.52\textwidth]{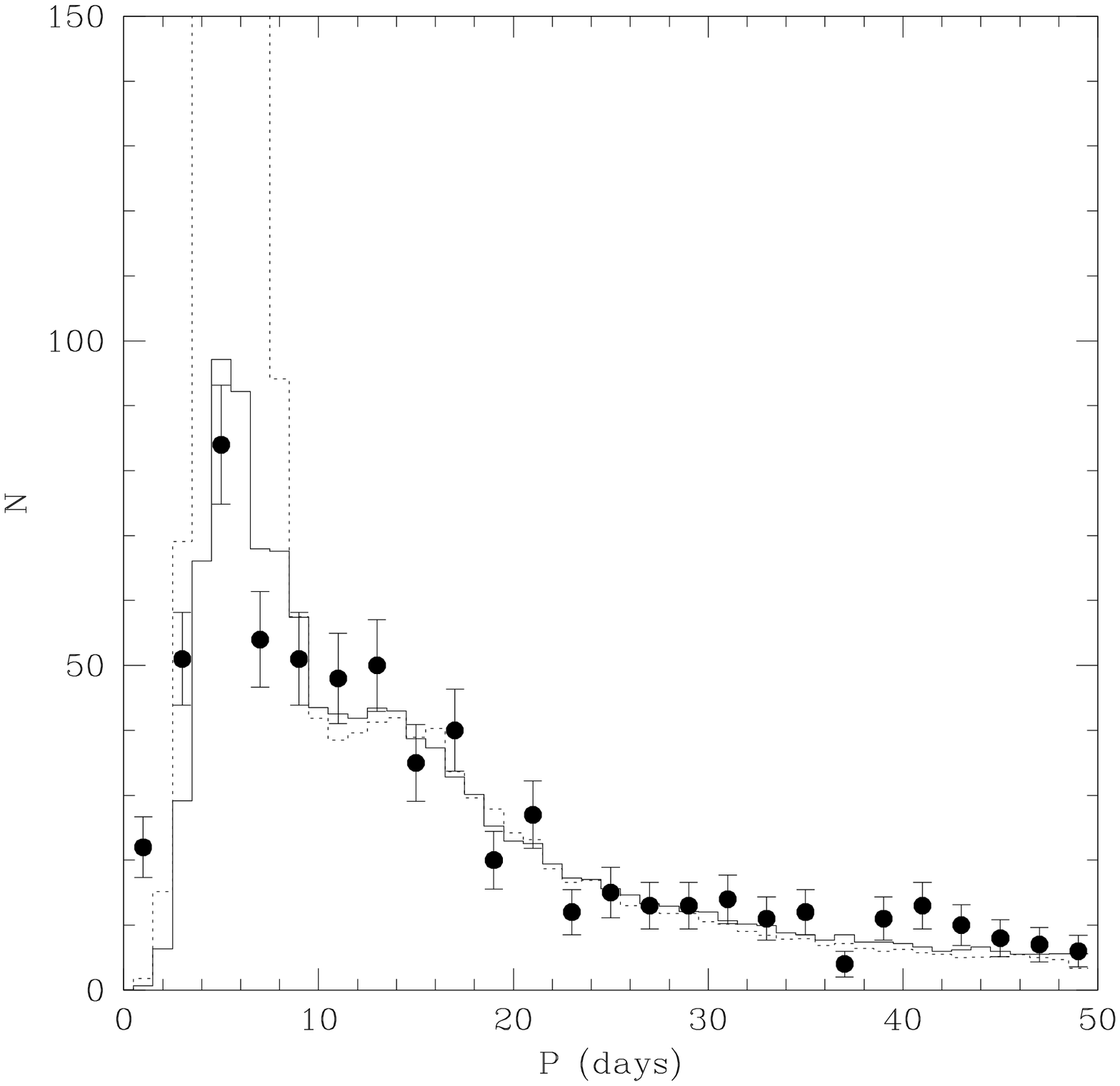},\includegraphics[width=0.52\textwidth]{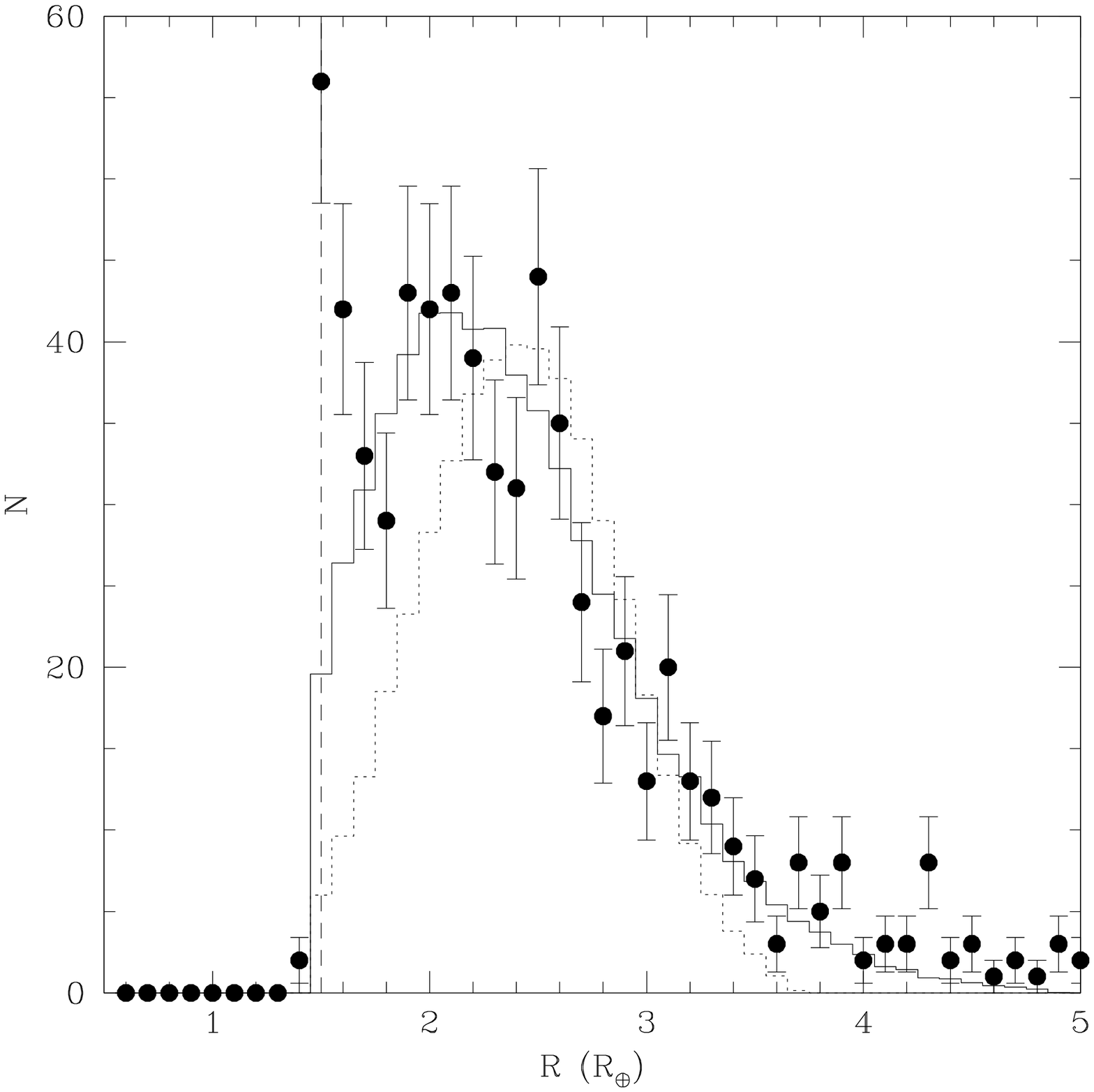}}
\caption{Solid points are the data, for all Kepler tranets, in both single and multiple systems, for the sample
defined in the text.
The solid histogram is for a model with $R'= 1.2$ , while the dotted histogram is the same, but in the case where we have
adjusted the inner edge of the distribution (see \S~\ref{Edge}).}
\label{Pdis}
\caption{The points show the distribution of radii for the planetary candidates in the observed sample, while the dashed
line indicates the cutoff radius. The dotted histogram is the distribution that emerges from our default model with $R'=1.2$.
It reproduces the average value of the observed distribution but clearly undercounts both the low and high radius tails. The
solid histogram shows the result if we use the mass dependant radius correction $R'=1 + 0.05 (M/M_{\oplus})$, which does a better
job of reproducing the observed shape.}
\label{Rbin}
\end{figure}

One mismatch observed in the previous subsection is the overproduction of tranets with orbital periods $<10$~days. This is due in large part
to the assumption that all protoplanetary disk have the same inner edge. To demonstrate this, 
 we consider a modification to our Monte Carlo model in which we adopt an
inner edge to the original disk of 0.08~AU, in 55\% of cases. In the other 45\% of cases, we adopt the original model. We use
the same mass, orbital spacing, eccentricity and inclination distributions as before. In effect, we simply remove those planets
that lie inside 0.08~AU in 55\% of the cases. We show the results of this model in Figure~\ref{Pdis}. This modification does not
change the quality of the fits to the $\xi$ distribution or the period ratio distribution.

Similarly, we have not attempted to directly match the observed distribution of radii, because of the distinct possibility that there
may be significant stochastic variation in the mass of hydrogen accreted/retained by planets. Nevertheless, one can match the observed
radius distribution with a suitable choice of mass-radius relation. Figure~\ref{Rbin} shows the observed distribution of tranetary
radii for the sample identified in \S~\ref{Obs}, compared to two distributions derived from our Monte Carlo model. The dotted histogram
shows the results for $R'=1.2$, which captures the median value of the population but does not match the shape. The solid histogram shows
the result from the same model except that we now adopt $R' = 1 + 0.05 (M/M_{\oplus})$. This provides a good fit to the tail of the
distribution at larger radii, and still matches the other observational constraints. This relation bears some resemblance to both the
interpolation of L11 and the linear fit of Wu \& Lithwick (2012a), although the nature of the fit means it is weighted more by lower
mass objects than the higher masses as in the case of Wu \& Lithwick. The effective physical model described by this relation is one
in which planets have Hydrogen mass fractions $< 1\%$ for masses $\sim M_{\oplus}$, increasing gradually to roughly 50\% at masses
$\sim 30 M_{\oplus}$.

\section{Discussion}
\label{Discuss}

Our study follows on the heels of several previous analyses of the Kepler multiple tranet candidate systems (Lissauer et al. 2011b; Fabrycky et al. 2012). All of these studies agree that the underlying planetary configuration is best modelled by a population that has
an inclination dispersion of only a few degrees, and that this is also consistent with the radial velocity sample (Lissauer et al. 2011b; Fabrycky et al. 2012; Figueira et al. 2011; Tremaine \& Dong 2012; Fang \& Margot 2012). Our analysis affirms these conclusions with the additional benefit of being drawn directly
from a physical model, rather than a set of ad hoc postulates. Our inability to match the single tranet numbers is also foreshadowed either
directly (L11) or by best fit models that require a large fraction of low multiplicity systems (Tremaine \& Dong 2012; Fang \& Margot 2012).

The fact that our underlying model always produces three or more final planets brings this issue into stark contrast. The most convenient explanation
is that that the rate of false positives is higher amongst the single tranet sample than the multiples, but this would require that 
current estimates of the false positive rate are severely in error. Lissauer et al. (2012) have argued that the rate of false positives amongst the
multiple systems is very low, but this necessarily excludes the single candidate systems.
Santerne et al. (2012) has claimed that the false positive rate for Jupiter-size candidates is higher ($\sim 35\%$)
than originally estimated (Morton \& Johnson 2011), but it is not clear that the same false positive rate will apply to systems with smaller eclipses.
Detailed simulations of the Kepler detection process suggests a false positive rate in the range 7\%-16\% for the size of planets in
our sample (Fressin et al. 2013). 
Thus, it seems that we require either an entirely separate contribution of single tranet systems or a process that modifies or edits
some fraction of the systems that assemble according to our model. 

\begin{figure}
\centerline{\includegraphics[width=0.52\textwidth]{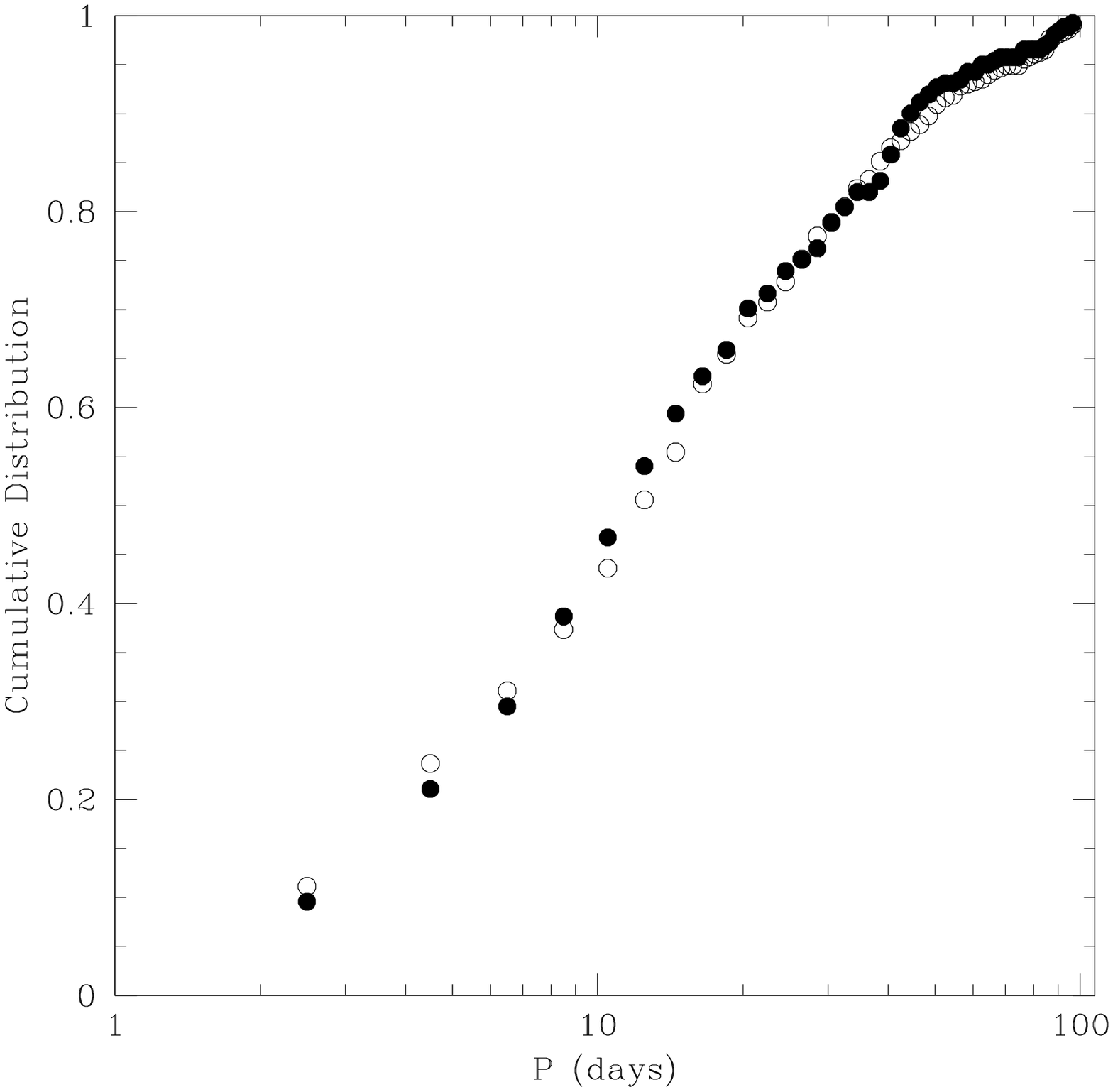},\includegraphics[width=0.52\textwidth]{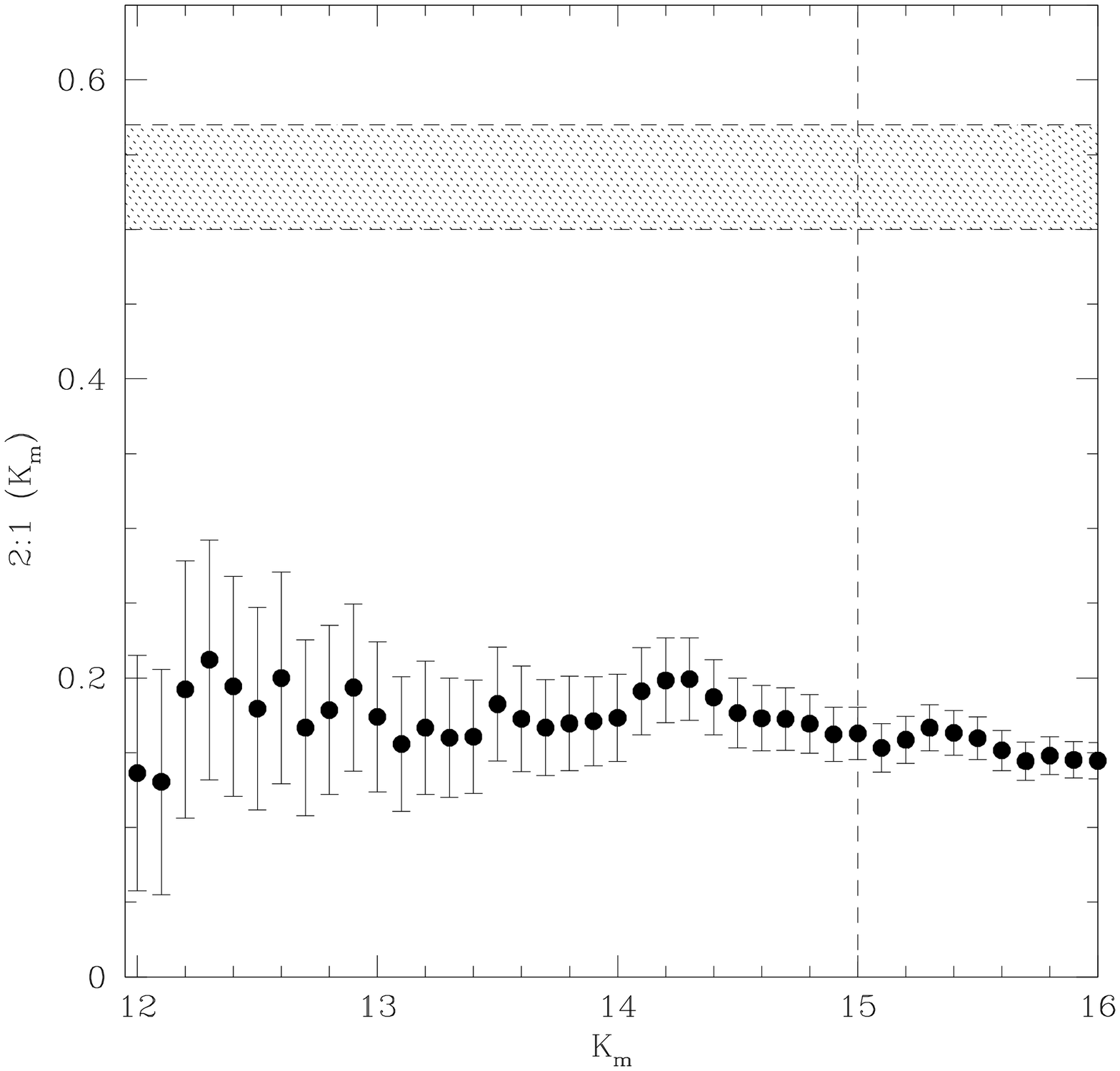}}
\caption{The solid points are the cumulative distribution with period for Kepler tranets
in multiple systems. The open points are the same quantity for single Kepler tranets. There is no
evidence that the two tranet samples are drawn from different distributions.}
\label{Pcomp}
\caption{The solid points show the ratio of double to single tranet systems, if we restrict our sample to
host stars brighter than Kepler magnitude $K_m$. Error bars are dictated by Poisson errors on the double tranet systems.
The vertical dashed line is the limit we use to define our sample in \S~\ref{Obs}, and the shaded region shows the range
of values we expect, based on our simulations. If some double tranet systems were being identified as only single tranet
systems because of low signal-to-noise, we would expect this ratio to decrease.}
\label{RatK}
\end{figure}

Although we have neglected any effects due to large-scale, post-assembly planetary migration in this calculation, it could potentially
be the alternative contributing process we seek, especially if the late-time instability of resonant chains (Terquem \& Papaloizou 2007; Ida \& Lin 2010) produces
primarily systems of low multiplicity. However, in that event, it is surprising that both processes produce essentially the
same orbital period distribution. 
 Figure~\ref{Pcomp} shows the cumulative distribution for tranets ($1.5 R_{\oplus} < R < 10 R_{\oplus}$) in single transit systems (open circles) and
multiple transit systems (filled circles). There is little evidence of any difference between the two populations that might indicate a different
origin scenario for $\sim 50\%$ of the single systems. Furthermore, there is also little evidence of any relative incompleteness based on
host star magnitude. Figure~\ref{RatK}
shows the ratio of double to single tranet systems as a function of the Kepler magnitude of the host stars. We see that the ratio remains essentially
constant from Kepler magnitudes 12--16, well beyond our nominal completeness limit.

 Alternatively, this could simply indicate that our model is incomplete in the sense that the systems
are assembled as we describe, but are then subjected to further dynamical influences that remove some planets from the system and
reduce the multiplicity. Indeed, we have simulated systems of terrestrial class planets only, without consideration of the interaction
with potential gas giant planets in the same system, despite the fact that we know such planets exist in our own solar system and others.
It is well known that even limited migration by giant planets during the loss of the protoplanetary nebula can cause the frequencies
of secular interactions to evolve, potentially resulting in resonant interactions, eccentricity excitation and eventual 
instability in lower mass planetary components (Ward 1981; Nagasawa, Lin \& Ida 2003; Levison \& Agnor 2003; Thommes, Nagasawa \& Lin 2008).
Alternatively, giant planet secular evolution by gravitational interactions alone may be enough to stimulate dynamical evolution
of lower mass systems (e.g. Veras \& Armitage 2003; Laskar 2008; Lithwick \& Wu 2011; Matsumura, Ida \& Nagasawa 2012)
Thus, it is possible that some of this dichotomy may simply reflect the role of a more distant component of giant planets in
some systems but not in others.

\subsection{W(h)ither Migration?}

We have demonstrated that a simple in situ assembly model can reproduce the broad distributions of tranetary periods
and inclinations observed in the Kepler database. However, this model seems to underpredict the frequency of close pairs (i.e. pairs
with period ratios $<2.5$) nor does it completely reproduce the observed structure near the 3:2 and 2:1 commensurabilities.
These suggest that, while the broad organisation of the observed planetary systems appears to be consistent with a
primarily gravitational process, matching the fine structure will likely require a model that includes some means of
 additional dissipation.

Part of this may be provided by tidal dissipation in the planets. We have already demonstrated that the fit
to the $\xi$ distribution can be improved if we allow for the circularisation of orbits inside $\sim 0.20$AU (corresponding
to $Q \sim 10$--50, depending on planetary mass and radius). This process is promising in that it can reduce the eccentricity
while preserving the inclination distribution of the planetary orbits, which seems to fit the observations well.

It is somewhat surprising that there is so little need for migration in fitting these observations. The rapid movement of
planetary mass objects in a gas disk has been a topic of great interest for the last several years. The strength of
this interaction was expected to result in very few planets of the observed mass range at the orbital periods observed, (e.g. Ida \& Lin 2008). One
way to slow the rate of migration is for strong gravitational interactions between planets to counteract the transfer of
angular momentum from orbit to gaseous disk, but this is expected to lead to systems with a large fraction of the systems
in or close to resonance (Terquem \& Papaloizou 2007; Ida \& Lin 2010). Although some resonant systems are seen, it appears as though a migration model requires a significant
additional stochastic forcing to match the observed period distribution (Rein 2012).

It is possible that this mismatch simply reflects the limitations of current models for how planets interact with the gaseous disk, as 
the effects of migration depend critically on the interaction between the planetary forcing and the structure of the disk
(e.g. Terquem 2003; Laughlin, Steinacker \& Adams 2004; Johnson, Goodman \& Menou 2006; Paardekooper \& Mellema 2006;
Paardekooper et al. 2010; Bromley \& Kenyon 2011; Hasegawa \& Pudritz 2011; Kretke \& Lin 2012). This is because the speed and even direction of
migration depends on the imperfect cancellation between the torques exerted by the disk interior and exterior to the planet.
One possibility, suggested by several authors (e.g. Masset et al. 2006), is that the torque may reverse sign at particular locations in the disk, leading to
`planet traps' -- preferred locations where planets assemble. It is possible that our power law surface density profile may simply represent an
averaged version of a disk with several such preferred locations, and that some disks may have more localised distributions which could
contribute to the overabundance of low multiplicity systems. Indeed, this might help to place our Solar System in the Kepler context, as our own
terrestrial planet system is potentially the outcome of a disk with a single localised planet trap (Hansen 2009).

\subsection{Predictions and Speculations}

The primary goal of this study is to demonstrate that a simple model of in situ assembly of rocky planets can
reproduce the distribution of observed parameters in the Kepler multiple tranet sample. However, given that
consistency, it is useful to attempt to outline additional analyses or observations that can further refine the
comparison. 

\begin{figure}
\centerline{\includegraphics[width=0.52\textwidth]{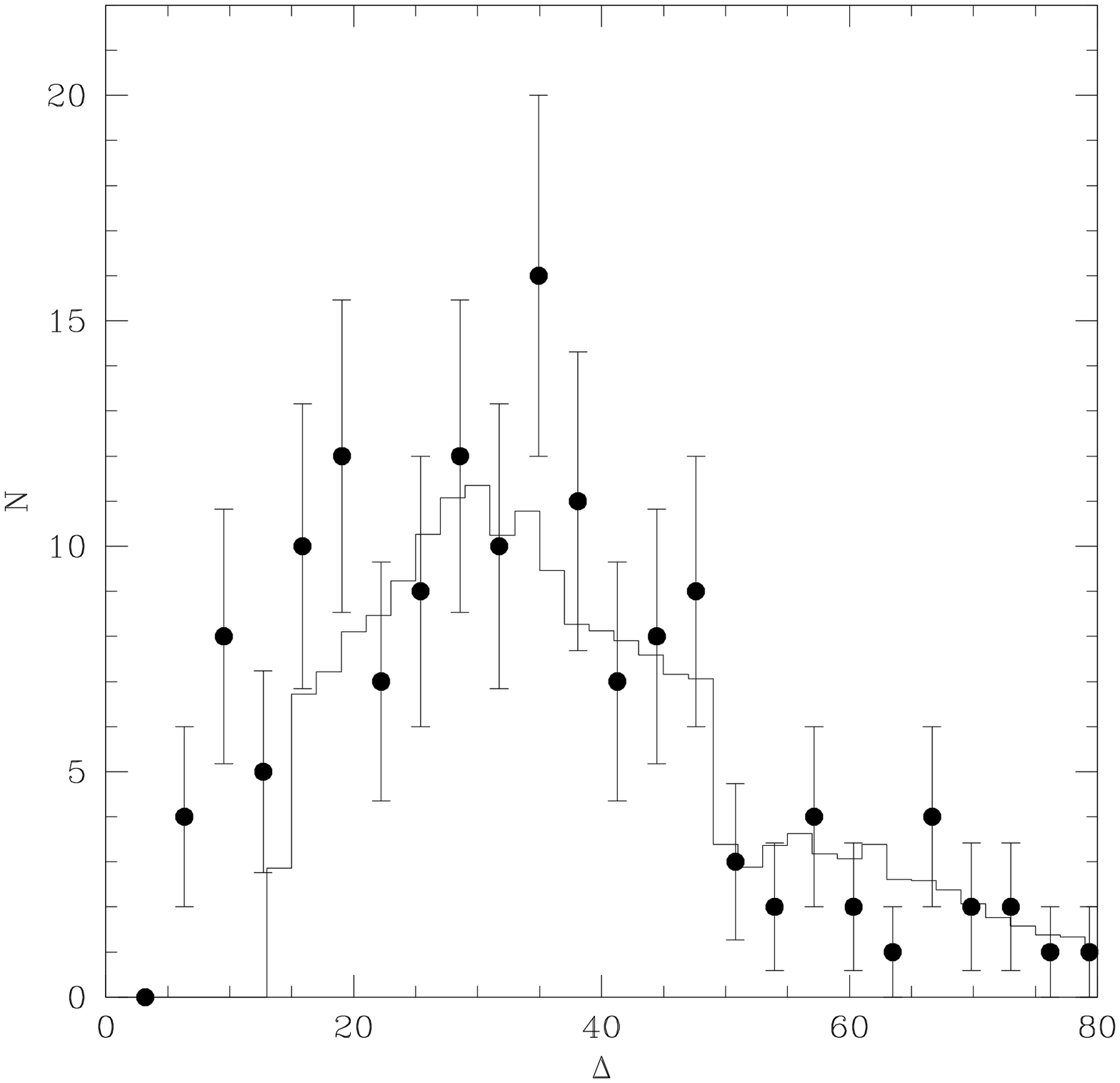},\includegraphics[width=0.52\textwidth]{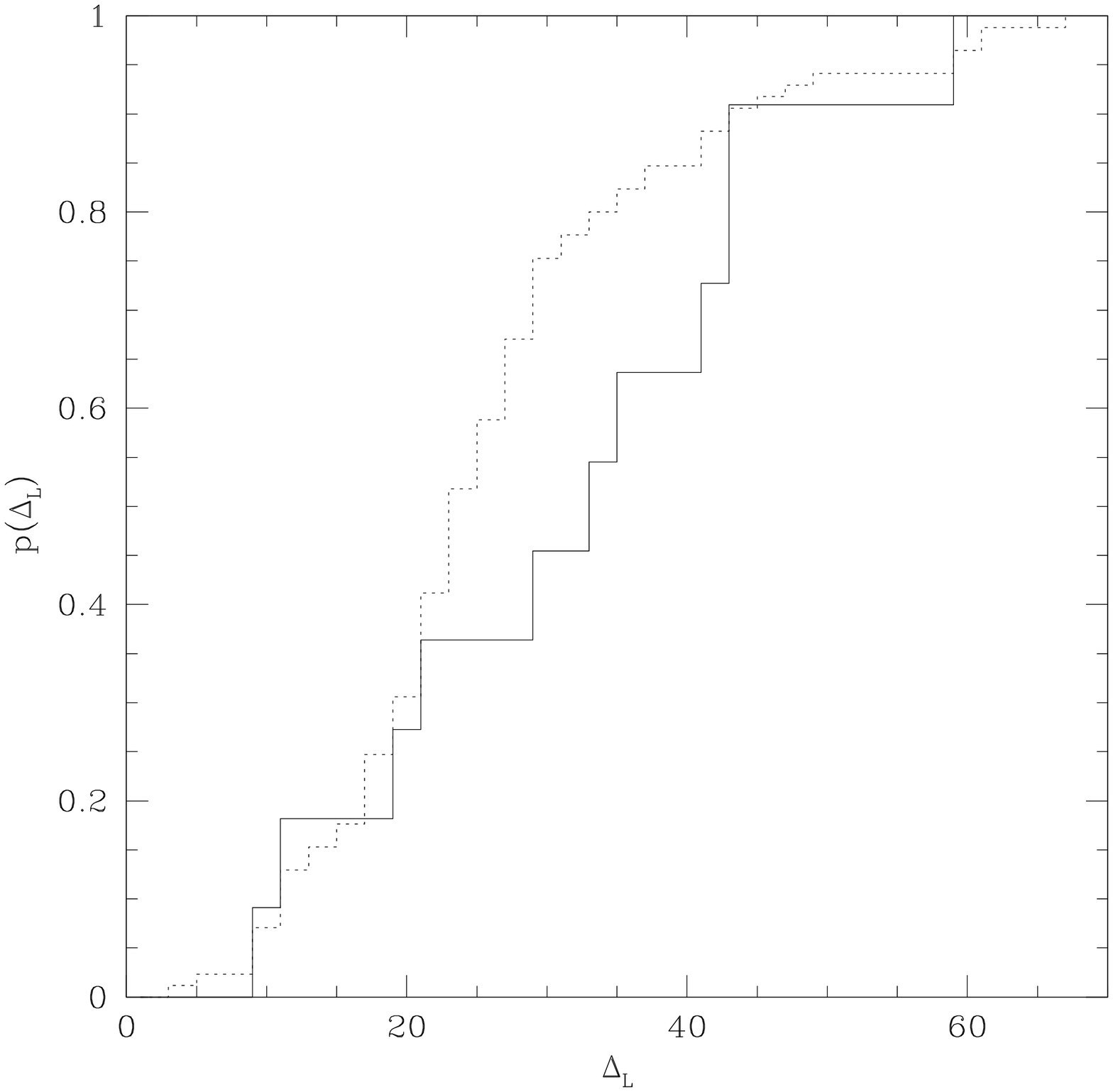}}
\caption{The filled circles are the distribution of $\Delta$ values for the observational sample as defined in the
text, for the Batalha et al. (2012) catalog. The values of $\Delta$ in each case were calculated using the mass-radius relation of L11, and then divided by a factor of
0.64. This was to bring the observed break into line with that observed at $\Delta \sim 50$ in the output from our Monte Carlo model (solid histogram).
Systems with values larger than this are very likely to harbour an undetected planet between the observed pair.}
\label{Ddis}
\caption{The solid histogram shows the cumulative distribution in $\Delta_L$ of the 11 candidate pair systems which
pass our stellar and planetary cuts and show TTVs of more than $1 \sigma$ statistical significance. The dotted histogram shows
the same but for the full sample of tranet pairs, selected in the same way except without taking account of TTV signatures.}
\label{TTV}
\end{figure}

Some of the consequences are rather obvious. The fact that neighbouring tranets with large period ratios 
can harbour additional unseen planets has been noted before (e.g. L11), but we can provide a prediction
of the quantitative threshold for such predictions. We have seen that our simulations rarely produce
neighbouring systems with $\Delta>50$, suggesting that multiple systems with larger values should harbor planets in these apparent gaps.
 Our model suggests planetary systems are factors of 2--3 less densely packed than a 
`maximally packed' system just on the edge of stability, so that the threshold of $\Delta \sim 50$ should
mark a change in the frequency of non-transiting companions. Indeed, Figure~\ref{Ddis} shows that the 
distribution of $\Delta$ that emerges from the simulations shows a qualitative break at $\Delta \sim 50$.
The high $\Delta$ tail of this distribution is thus the group that should contain 
non-transiting interlopers. We can compare this to the observed distribution of $\Delta$ in the observed
sample. In this case we use the catalog of Batalha et al. (2012) in order to use later additions to the catalog
to assess any predictive power our model comparison may uncover.
 A comparison with the data also requires the adoption of a mass-radius relation.
Using the mass-radius relation used by L11 and F12, to maintain consistency with previous analyses,
we define our observed $\Delta$ as $\Delta_L$. The resulting observed distribution shows a very similar
feature, but the break occurs at $\Delta_L \sim 30$. 
 The similarity in shape between
the two distributions suggests a measure of renormalisation is in order, and that the values of $\Delta_L$ used
in L11 and F12 are likely underestimated by $\sim 60$\%, or that the L11 mass-radius relation overestimates
the planetary mass, at least in the lower mass regime. Table~\ref{DeltaTab} lists the observed pairs that
have $\Delta_L > 30$ in Figure~\ref{Ddis}. 
 If we estimate masses from orbital periods for the entries in this table with $30<\Delta_L<33$, using
equation~(\ref{Mtrend}), we find a range $38.4 < \Delta < 50.5$ for 11 systems, with a median of $\Delta = 48$.
Thus, it suggests that the renormalisation from $\Delta_L=30$ to $\Delta = 50$ is consistent with the difference
in the adopted models for the planetary masses.

As can be seen from the notes to Table~\ref{DeltaTab}, there are several systems in which we have strong evidence
for the presence of planets in the ostensible gaps. The systems plotted in Figure~\ref{Ddis} were selected for
consistency with our model comparison, namely we restricted our attention to planets with radii $>1.5 R_{\oplus}$.
In two of the cases, additional tranets (KOI~70.05 and KOI~117.04) were already known but not included in the above comparison
because their radii fell below the radius cutoff. The addition of further data beyond the first six quarters 
 has also resulted in the detection of tranets in the gaps in three other (KOI-116, KOI-593 and
KOI-671) systems. In three systems (KOI-316, KOI-341 and KOI-582) additional observations have cast doubt on either the
validity of one of the candidates themselves, or on one of the periods (which would therefore change the $\Delta_L$).

In addition to directly observing a transit by an additional planet candidate, it is possible
 to identify the presence of non-transiting companions by observing variations in
the timing of transits by known tranets, due to the gravitational perturbation of the unseen companion (Agol et al. 2005; 
Holman \& Murray 2005; Veras, Ford \& Payne 2011).
Ford et al. (2012) identified 1436 systems with potential transit timing variations (TTVs). If we restrict our
attention to systems with two candidate tranets, with the same stellar, planetary and magnitude cuts as
before, we are left with 11 potential systems with median deviations from a linear ephemeris that are 
larger than the quoted dispersions representing transit timing accuracy. Figure~\ref{TTV} shows the cumulative distribution of these 11 systems
as a function of $\Delta_L$, along with the same distribution for the full sample of observed double candidate
systems (subject to the same cuts). We see that the systems with potential TTVs are indeed skewed to larger
$\Delta_L$, and that the maximum deviation occurs at $\Delta_L \sim 30$, as expected. The small number of
candidate systems means that this result is still statistically weak (A Kolmogorov-Smirnov test suggests that
the TTV-selected distribution could be drawn from the larger sample with a probability of 7\%), but this should hopefully
improve with longer baselines of observation. Furthermore, Steffen et al. (2012) identify candidate pairs which may show
anticorrelated TTV signatures, and many of the pairs in Table~\ref{DeltaTab} are identified as worthy of further study.
A better comparison may be possible soon as Mazeh et al. (2013) promise an improved catalog will be shortly forthcoming, based on a longer baseline of data.

It is also worth noting that a first prediction of this model, based on multiplicity statistics,
 was made in HM11. In that case, three multiple planet systems were identified in which the detection
of additional planets would improve the match with the models. The recent announcement of additional
planets in one of these systems, HD40307 (Tuomi et al. 2012), does indeed improve the agreement with
the original models. Although this is hardly a revolutionary prediction (a linear trend in the
radial velocity data was already 
known at the time of writing), it is still encouraging that additional detections improve the agreement
between the model and the data.

If the overabundance of single tranet systems is indeed the result of a systemic instability induced by
the presence of a giant planet system at larger radii, then it would of interest to study two samples
(one set of multiple candidate systems and another of single candidates of similar radius) of Kepler systems with radial velocities,
in order to characterise whether there are any significant differences which may betray divergent dynamical histories.
One possible signature would be if the single short-period candidate systems showed two or more giant planets, while
multiple short-period candidate systems showed only zero or one giant planet. Such a difference might point the way
to secular evolution and resonance playing a role in the disruption of inner planetary systems. 

Improved eccentricity constraints would also enable better constraints on the model. As showed above, the fit to the $\xi$
distribution (Figure~\ref{xiplot}) got better for larger $a_{circ}$ (i.e. stronger tidal damping in the
planets).  The current constraints on the eccentricities due to transit durations (Moorhead et al. 2011) find
little evidence for circularisation. However, their contraint is weakened by uncertainty in the stellar
parameters and strictly only applies for cooler stars than our nominal host.
 Wu \& Lithwick (2012a) find that a significant
fraction of their systems with strong TTVs show very small free eccentricities, consistent with a level
of tidal dissipation similar to what we have assumed here (although there appears to be a minority of
systems with significant eccentricities, which are harder to explain). 

\section{Conclusions}

We have studied the assembly of planetary embryos into final planets for 
a disk of total mass $20 M_{\oplus}$ (interior to 1 AU) around a solar mass star.
We quantify the distribution of final masses, eccentricities and inclinations, as well as the spacings
of the final planets, and use these to generate a large population of simulated systems in a Monte Carlo
model. The output from these simulations is used to quantify their detectability in a transit survey
such as that conducted by the Kepler mission. 

Our principal conclusions are
\begin{itemize}
\item The relative frequency of multiple transiting systems is well matched by the simulations. However,
our model underpredicts the frequency of single tranet systems. If this is not due to unquantified selection effects, then
it suggests that either an independent process
produces mostly low multiplicity systems, or that some fraction of the planetary systems formed by our model undergo
additional perturbations which reduce their multiplicity.\\
\item The distribution of mutual inclinations, quantified by the velocity-normalised transit duration ratio $\xi$ (F12),
is well matched by the simulations, especially if we allow for tidal damping of eccentricity for short period orbits. This is
encouraging as it provides a physical basis for the observed inclination distribution of the Kepler tranet candidate
sample. \\
\item Our default model matches the distribution of periods well for periods $>10$~days, but overpredicts the number of short
period tranets. This deficit can be corrected by allowing for a variation in the inner edge of the initial disk, as demonstrated
in \S~\ref{Edge}. \\
\item Our model underpredicts the number of close tranet pairs. This is the one observable that suggests that some
extra (non-tidal) dissipation is required during the planetary assembly. At present there appear to be at least two possible ways to achieve this.
One would be to cause a moderate amount of radial migration to bring pairs into closer alignment. Alternatively, it appears as though
the observed period ratio distribution would be matched better if the relative fraction of high multiplicity to low multiplicity systems 
were increased in the Monte Carlo model. This could potentially be achieved with a moderate amount of inclination damping by interactions
with a protoplanetary gas disk. However, too much damping would destroy the agreement
 with the $\xi$ distribution, so it might be possible to place some interesting constraints on the conditions during planet assembly.\\
\item We suggest that most observed tranet pairs with $\Delta_L>30$ contain additional non-transiting planets between them,
and we find weak statistical evidence (limited by currently small sample sizes) that pairs with non-negligible TTV signatures are preferentially found with values
above this threshold. Larger and improved data sets can help to refine this test of the model.
\end{itemize}

Our assumption of a universal power law disk of fixed mass, which assembles by purely gravitational interactions, is clearly an 
oversimplification of what is expected to be a very complex physical problem. However, the broad distributions of inclination,
multiplicity and planetary spacing match well to both observations directly and to prior inferences based on ad hoc models. The
fact that our model has a physical basis places some of the issues into starker contrast. In particular, the observed excess of
single (or low multiplicity) tranetary systems suggests that additional processes are needed to provide a complete model. 
The fact that a purely
gravitational process produces a final distribution that matches the observations places constraints on the amount of dissipation
expected from interactions with the protoplanetary disk. Some level of dissipation is required to match the period ratio distribution,
but it needs to be weak enough to destroy neither the broad spacing in semi-major axis nor the inclination distribution.

Finally, our ability to fruitfully perform such analyses illustrates the value of large scale, homogeneously selected observational datasets.
The requirement that our models match a broad range of observed systems promises to finally allow us to begin to place meaningful
constraints on the mechanisms by which planetary systems form and assemble.

\acknowledgements 

N.M. is supported in part by NSERC of Canada and by the Canada Research Chair program. B.H. thanks the group
at NASA Ames for discussions that stimulated the discussion in \S~\ref{Assemble} and both express their enthusiasm for the work done
by the Kepler team in producing this wonderful data set.

\newpage

\begin{deluxetable}{lcccccc|lcccccc}
\tablecolumns{14}
\tablewidth{0pc}
\tablecaption{Final States of Assembly Simulations
\label{OutTab}}
\tablehead{ $\#$ & 
\colhead{N} & \colhead{$M_{big}$} & \colhead{$S_d$}  & \colhead{$<a>_M$} & \colhead{S$_s$} & \colhead{S$_c$} &
$\#$ &
\colhead{N} & \colhead{$M_{big}$}  & \colhead{$S_d$}  & \colhead{$<a>_M$}  & \colhead{S$_s$} & \colhead{S$_c$} \\
 &   &\colhead{($M_{\oplus}$)} & \colhead{($\times 10^{-2}$)} & \colhead{(AU)} & & & & \colhead{($M_{\oplus}$)} & \colhead{($\times 10^{-2}$)} &
 \colhead{(AU)} &\\
 }
\startdata
  1 &  6 &    7.8 &   0.61 & 0.46     & 21.4     & 7.66     &  51 &  3 &    8.7 &   2.03 & 0.29     & 42.0     & 9.25     \\
   
  2 &  3 &    5.7 &   1.50 & 0.26     & 35.4     & 11.8     &  52 &  4 &    7.1 &   2.62 & 0.37     & 32.4     & 6.90     \\
   
  3 &  5 &    3.5 &   0.57 & 0.30     & 24.5     & 8.28     & 53 &  5 &    5.3 &   0.60 & 0.47     & 24.5     & 5.78     \\
   
  4 &  7 &    3.4 &   0.18 & 0.27     & 18.7     & 8.97     & 54 &  4 &    5.2 &   1.64 & 0.32     & 29.8     & 7.72     \\ 
   
  5 &  6 &    5.6 &   0.74 & 0.38     & 21.9     & 7.84     & 55 &  6 &    5.7 &   1.06 & 0.36     & 22.0     & 7.88     \\
   
  6 & 10 &    2.6 &   0.08 & 0.27     & 10.3     & 12.1     & 56 &  4 &    6.9 &   0.45 & 0.35     & 30.1     & 7.91     \\
   
  7 &  6 &    6.1 &   0.45 & 0.44     & 20.8     & 7.49     & 57 &  4 &    5.1 &   0.84 & 0.30     & 29.4     & 8.10     \\
   
  8 &  5 &    4.2 &   1.54 & 0.38     & 25.3     & 7.40    & 58 &  4 &    7.2 &   3.86 & 0.37     & 31.6     & 8.52     \\
   
  9 & 10 &    2.9 &   0.08 & 0.39     & 12.3     & 11.0    & 59 &  4 &    8.5 &   1.51 & 0.41     & 31.2     & 8.13     \\
   
 10 &  4 &    6.5 &   2.44 & 0.29     & 29.7     & 11.5    & 60 &  5 &    5.3 &   1.90 & 0.43     & 24.7     & 7.82     \\
   
 11 &  7 &    4.3 &   0.55 & 0.47     & 18.0     & 7.55    & 61 &  6 &    5.9 &   0.60 & 0.40     & 21.1     & 7.07     \\
   
 12 &  7 &    3.8 &   0.26 & 0.31     & 18.3     & 8.68    & 62 &  3 &    5.9 &   1.63 & 0.31     & 37.0     & 10.1     \\
   
 13 &  4 &    7.2 &   2.81 & 0.46     & 30.6     & 8.04    & 63 &  4 &    5.5 &   2.52 & 0.33     & 30.2     & 7.62     \\
   
 14 &  5 &    5.1 &   1.61 & 0.39     & 23.5     & 8.86    & 64 &  5 &    6.6 &   1.18 & 0.33     & 24.9     & 10.3     \\ 
   
 15 &  3 &    8.5 &   1.98 & 0.35     & 41.3     & 9.17    & 65 &  4 &    5.3 &   1.45 & 0.33     & 29.5     & 10.5     \\
   
 16 &  7 &    3.2 &   0.59 & 0.26     & 19.5     & 9.08    & 66 &  5 &    8.8 &   1.54 & 0.35     & 25.4     & 8.93     \\
   
 17 &  7 &    4.2 &   0.35 & 0.36     & 18.5     & 8.43    & 67 &  3 &    8.6 &   4.02 & 0.24     & 38.4     & 11.5     \\
   
 18 &  5 &    5.4 &   1.93 & 0.45     & 25.3     & 7.44    &  68 &  4 &    9.2 &   1.48 & 0.33     & 31.2     & 10.6     \\
   
 19 &  5 &    4.6 &   0.30 & 0.31     & 26.5     & 8.60    & 69 &  5 &    6.8 &   0.51 & 0.44     & 24.6     & 7.52     \\ 
   
 20 &  3 &    7.2 &   3.34 & 0.35     & 39.7     & 8.97    & 70 &  4 &    4.7 &   4.46 & 0.40     & 29.1     & 7.81     \\
   
 21 &  6 &    4.4 &   2.51 & 0.37     & 21.4     & 8.37    & 71 &  5 &    5.7 &   0.75 & 0.38     & 24.7     & 7.53     \\
   
 22 &  7 &    4.6 &   0.47 & 0.35     & 18.9     & 8.32    & 72 &  4 &    8.5 &   4.60 & 0.37     & 31.6     & 7.65     \\
   
 23 &  5 &    7.4 &   1.53 & 0.35     & 24.5     & 10.2    & 73 &  8 &    4.2 &   0.67 & 0.45     & 16.5     & 6.27     \\
   
 24 &  5 &    5.2 &   0.53 & 0.36     & 23.6     & 9.00    & 74 &  5 &    5.4 &   1.38 & 0.40     & 24.8     & 6.64     \\
   
 25 &  4 &    6.4 &   0.60 & 0.43     & 31.2     & 7.14    & 75 &  7 &    3.8 &   0.79 & 0.44     & 18.2     & 6.35     \\
   
 26 &  5 &    5.5 &   0.48 & 0.34     & 24.7     & 8.26    & 76 &  4 &    6.8 &   3.48 & 0.32     & 31.0     & 8.78     \\ 
   
 27 &  5 &    5.0 &   1.62 & 0.38     & 24.7     & 8.14    & 77 &  4 &    7.3 &   5.64 & 0.30     & 33.0     & 8.46     \\
   
 28 &  6 &    4.3 &   0.35 & 0.28     & 22.5     & 8.99    & 78 &  5 &    7.0 &   1.13 & 0.43     & 25.0     & 6.81     \\
   
 29 &  6 &    3.5 &   0.30 & 0.29     & 21.1     & 8.62    & 79 &  6 &    4.7 &   0.50 & 0.42     & 21.0     & 7.16     \\
   
 30 &  7 &    5.1 &   0.83 & 0.41     & 18.4     & 7.86    & 80 &  4 &    6.3 &   1.30 & 0.36     & 30.5     & 8.28     \\
   
 31 &  6 &    3.9 &   0.41 & 0.26     & 22.1     & 9.03    & 81 &  4 &    5.9 &   5.81 & 0.41     & 31.6     & 7.48     \\
   
 32 &  9 &    2.2 &   0.15 & 0.32     & 15.0     & 8.85    & 82 &  7 &    4.2 &   0.23 & 0.41     & 17.8     & 8.23     \\
   
 33 &  5 &    4.6 &   0.30 & 0.32     & 25.9     & 8.44    & 83 &  5 &    7.2 &   0.59 & 0.37     & 23.3     & 7.91     \\
 
 34 &  8 &    4.8 &   0.24 & 0.29     & 16.9     & 8.98    & 84 &  5 &    5.5 &   0.56 & 0.40     & 25.4     & 7.67     \\
   
 35 &  4 &    5.7 &   1.56 & 0.38     & 30.8     & 7.19    & 85 &  6 &    5.8 &   0.23 & 0.46     & 20.7     & 6.32     \\
   
 36 &  5 &    5.9 &   0.88 & 0.33     & 24.7     & 9.31    & 86 &  4 &    7.0 &   0.60 & 0.39     & 30.0     & 7.98     \\
   
 37 &  7 &    3.2 &   0.36 & 0.34     & 18.4     & 8.42    & 87 &  5 &    5.8 &   1.90 & 0.44     & 25.4     & 6.56     \\
   
 38 &  5 &    6.9 &   2.63 & 0.37     & 25.5     & 8.82    & 88 &  4 &    8.0 &   2.96 & 0.53     & 30.3     & 5.31     \\
   
 39 &  4 &    6.0 &   1.12 & 0.36     & 29.5     & 8.06    & 89 &  3 &    5.7 &   3.35 & 0.44     & 35.8     & 11.2     \\
   
 40 &  6 &    3.8 &   0.40 & 0.32     & 21.0     & 8.07    & 90 &  4 &    8.6 &   3.78 & 0.37     & 29.6     & 9.19     \\
   
 41 &  5 &    7.9 &   0.97 & 0.38     & 23.2     & 10.5    & 91 &  6 &    4.6 &   0.68 & 0.41     & 20.2     & 8.50     \\
   
 42 &  5 &    4.7 &   1.10 & 0.42     & 24.4     & 6.42    & 92 &  4 &    6.4 &   1.31 & 0.38     & 30.5     & 8.42     \\
   
 43 &  6 &    5.9 &   0.68 & 0.39     & 21.6     & 7.92    & 93 &  4 &    6.8 &   1.47 & 0.41     & 29.5     & 7.40     \\
   
 44 &  4 &    8.6 &   0.74 & 0.39     & 32.7     & 7.51    & 94 &  4 &    5.8 &   1.23 & 0.38     & 29.3     & 7.92      \\
   
 45 &  5 &    3.8 &   0.91 & 0.38     & 24.2     & 8.15    & 95 &  5 &    6.2 &   1.47 & 0.50     & 23.3     & 6.64      \\
   
 46 &  8 &    7.1 &   0.24 & 0.44     & 16.2     & 7.94    & 96 &  4 &    7.5 &   2.14 & 0.33     & 29.9     & 8.64     \\
   
 47 &  4 &    7.1 &   2.68 & 0.44     & 30.3     & 6.02    & 97 &  3 &   10.3 &   1.32 & 0.30     & 39.1     & 8.85     \\
   
 48 &  4 &    7.6 &   1.69 & 0.43     & 30.6     & 6.93    & 98 &  5 &    6.8 &   1.17 & 0.45     & 25.2     & 6.27     \\ 
   
 49 &  8 &    4.1 &   0.36 & 0.45     & 16.2     & 6.52    & 99 &  5 &    5.2 &   1.34 & 0.42     & 25.0     & 6.33     \\
   
 50 &  7 &    4.9 &   0.19 & 0.45     & 18.2     & 7.67    & 100 &  7 &    4.9 &   0.47 & 0.46     & 18.2     & 7.94     \\
\end{deluxetable}

\begin{deluxetable}{ccccccc}
\tablecolumns{7}
\tablewidth{14cm}
\tablecaption{Candidates for Missing Planet Systems.
\label{DeltaTab}}
\tablehead{
\colhead{KOI pair} & \colhead{$P_{in}$} & \colhead{$P_{out}$} &
\colhead{$R_{in}$} & \colhead{$R_{out}$} & \colhead{$\Delta_L$} & \colhead{Notes}  \\
 & \colhead{(days)} & \colhead{(days)} & \colhead{($R_{\oplus}$)} & \colhead{($R_{\oplus}$)}
 }
\startdata
 70.01--70.03 & 10.85 & 77.61 & 3.1 & 2.8 & 43.6  & \tablenotemark{1} \\  
 108.01--108.02 & 15.97 & 179.6 & 2.94 & 4.45 & 42.6 &  \\ 
 116.01--116.02 & 13.57 & 43.84 & 2.58 & 2.68 & 30.4 & \tablenotemark{2}\tablenotemark{*} \\ 
 117.02--117.01 & 4.90 & 14.75 & 1.70 & 2.93 & 30.6 & \tablenotemark{3}\tablenotemark{*} \\  
 123.01--123.02 & 6.48 & 21.22 & 2.64 & 2.71 & 30.4 &  $^*$\\
 148.02--148.03 & 9.67 & 42.90 & 3.14 & 2.35 & 36.2 & \tablenotemark{4}\tablenotemark{*} \\
 150.01--150.02 & 8.41 & 28.57 & 2.63 & 2.73 & 31.2 & $^*$\\   
 223.01--223.02 & 3.18 & 41.0 & 2.52 & 2.27 & 60.3 & \\ 
 270.01--270.02 & 12.58 & 33.67 & 1.80 & 2.13 & 31.6 & \tablenotemark{5}\tablenotemark{*}\\
 275.01--275.02 & 15.79 & 82.20 & 1.95 & 2.04 & 49.4 & \\
 291.01--291.02 & 8.13 & 31.52 & 2.14 & 2.73 & 36.3 & \\
 316.01--316.02 & 15.77 & 157.06 & 2.72 & 2.94 & 50.1 & \tablenotemark{6} \\ 
 339.01--339.02 & 1.98 & 12.83 & 1.55 & 1.66 & 63.4 & $^*$\\
 341.01--341.02 & 4.70 & 14.34 & 1.58 & 2.98 & 31.0 & \tablenotemark{7}\\
 343.01--343.03 & 4.76 & 41.81 & 2.68 & 1.58 & 57.2 & $^*$\\
 377.03--377.01 & 1.59& 19.27 & 1.67 & 8.28 & 31.5 & \tablenotemark{8} \\ 
 416.01--416.02 & 38.1 & 18.21 & 2.81 & 2.71 & 88.3 & $^*$ \\
 438.01--438.02 & 5.93 & 52.66 & 1.74 & 2.10 & 62.9 & $^*$\\
 440.01--440.02 & 4.97 & 15.91 & 2.36 & 2.26 & 33.0 & \\
 448.01--448.02 & 10.14 & 43.61& 1.77 & 2.31 & 43.6 & \tablenotemark{9}\tablenotemark{*} \\  
 456.01--456.02 & 4.31 & 13.70 & 1.50 & 3.27 & 30.7 & \tablenotemark{10}\\
 464.01--464.02 & 5.35 & 58.36 & 2.63 & 6.73 & 34.2 & $^*$\\
 481.02--481.01 & 1.55 & 7.65 & 1.54 & 2.37 & 48.0 & \\
 481.01--481.03 & 7.65 & 34.26 & 2.37 & 2.44 & 40.1 & $^*$ \\
 497.01--497.02 & 4.43 & 13.19 & 1.53 & 2.74 & 32.0 & \\
 518.01--518.02 & 13.98 & 44.0 & 2.11 & 1.54 & 38.0 & \tablenotemark{11}\tablenotemark{*} \\ 
 528.03--528.02 & 20.55 & 96.67 & 3.12 & 3.27 & 33.9 & \\
 582.01--582.02 & 5.95 & 17.74 & 2.24 & 2.12 & 32.5 & \tablenotemark{12}\tablenotemark{*} \\ 
 593.01--593.02 & 10.00 & 90.41 & 2.63 & 3.79 & 44.1 & \tablenotemark{13} \\
 657.01--657.02 & 4.07 & 16.28 & 1.63 & 2.08 & 44.6 & $^*$\\
 671.01--671.03 & 4.23 & 16.26 & 1.66 & 1.61 & 47.9 & \tablenotemark{14} \\ 
 701.01--701.03 & 18.16 & 122.39 & 1.95 & 1.57 & 60.3 &  \\
 711.02--711.01 & 3.62 & 44.70 & 1.50 & 3.18 & 58.0 & \\
 718.01--718.02 & 4.59 & 22.71 & 2.57 & 3.06 & 38.0 & $^*$\\
 986.01--986.02 & 8.19 & 76.05 & 1.85 & 1.54 & 69.5 & $^*$\\
 1435.01--1435.02 & 10.45 & 40.72 & 1.93 & 2.04 & 42.1 & \\
 1945.01--1945.02 & 17.12 & 62.14 & 1.8 & 2.94 & 34.8 & \\   
 1952.01--1952.02 & 8.01 & 27.67 & 1.83 & 2.02 & 39.6 & \\   
\enddata
\tablenotetext{1}{KOI~70.05 is a known candidate from Q1--Q6 data, lies in this gap but was not included in the analysis
because $R_p=1.03 R_{\oplus}$.}
\tablenotetext{2}{Inclusion of Q1--Q8 data results in the discovery of KOI~116.04 in this gap.}
\tablenotetext{3}{KOI~117.04 is a known candidate from Q1--Q6 data and lies in this gap, but was not included in the analysis
because $R_p=1.07 R_{\oplus}$.}
\tablenotetext{4}{Reported as having a significant TTV signature in Mazeh et al. 2013}
\tablenotetext{5}{Reported as having a significant TTV signature in Mazeh et al. 2013}
\tablenotetext{6}{Inclusion of Q1--Q8 data appears to render the KOI~316.01 period less certain.}
\tablenotetext{7}{Mazeh et al. (2013) report a significant TTV signature for 341.01 and also suggest that the
real period may be half that reported. This would reduce the spacing and move the system below the $\Delta_L=30$ threshold.}
\tablenotetext{8}{Candidate KOI~377.01 shows a strong TTV signal in the Ford et al. (2012) compilation, but this
is most likely due to the proximity of the 377.01--377.02 pair to the 2:1 commensurability.}
\tablenotetext{9}{KOI~448.02 shows a possible TTV signature in the Ford et al. (2012) compilation, and is also reported
as having a significant TTV signature by Mazeh et al. (2013)}
\tablenotetext{10}{Reported as having a significant TTV signature in Mazeh et al. 2013}
\tablenotetext{11}{Inclusion of Q1--Q8 data adds a third, outer, candidate to the system, also with a 
large separation. This system could potentially hold two extra planets.}
\tablenotetext{12}{Further analysis suggests that KOI~582.01 is a false positive. Discovery of a new candidate 582.03 results
in a pair with $\Delta_L=21$.}
\tablenotetext{13}{KOI~593.02 shows a potential TTV signature in the Ford et al. (2012) compilation. Inclusion of Q1--Q8
data results in an additional candidate, KOI~582.03, in the gap.}
\tablenotetext{14}{Inclusion of Q1--Q8 data actually reveals two new tranets in this gap, while also reducing the radii of
all four tranets below $1.5 R_{\oplus}$.}
\tablenotetext{*}{This pair was noted to have a possible anticorrelated TTV signature in Steffen et al. (2012). ($\Xi >1$)}
\end{deluxetable}


\begin{references}
\reference{ASSC} Agol, E., Steffen, J., Sari, R. \& Clarkson, W., 2005, MNRAS, 359, 567
\reference{B04} Baraffe, I. et al., 2004, A\&A, 419, L13
\reference{B11} Batalha, N. et al., 2011, ApJ, 729, 27
\reference{B12} Batalha, N. et al., 2012, arXiv:1202.5852
\reference{BM} Batygin, K. \& Morbidelli, A., 2013, AJ, 145, 1
\reference{B06} Beaulieu, J. P. et al., 2006, Nature, 439, 437
\reference{Bo12} Boisse, I., 2012, A\&A, 545, A55
\reference{B10} Borucki, W. J. et al. 2010, ApJ, 713, L126
\reference{BK} Bromley, B. \& Kenyon, S. J., 2011, ApJ, 735, 29
\reference{BM96} Butler, R. P. \& Marcy, G. W., 1996, ApJ, 464, L153
\reference{BM97} Butler, R. P., Marcy, G. W., Williams, E., Hauser, H. \& Shirts, P., 1997, ApJ, 424, L115
\reference{C99} Chambers, J. E., 1999, MNRAS, 304, 793
\reference{C01} Chambers, J. E., 2001, Icarus, 152, 205
\reference{C09} Charbonneau, D., et al., 2009, Nature, 462, 891
\reference{CL} Chiang, E.  \& Laughlin, G., 2012, arXiv:1211.1673
\reference{ED11} Ehrenreich, D. \& Desert, J.-M., 2011, A\&A, 529, A136
\reference{FT} Fabrycky, D.  \& Tremaine, S., 2007, ApJ, 669, 1298
\reference{Fab} Fabrycky, D., et al., 2012, arXiv:1202.6328
\reference{F12} Ford, E. B. et al., 2012, ApJ, 756, 185
\reference{F13} Fressin, F., et al., 2013, arXiv:1301.0842
\reference{G08} Gaudi, B. S., et al., 2008, ApJ, 677, 1268
\reference{G93} Gladman, B., 1993, Icarus, 106, 247
\reference{GT80} Goldreich, P. \& Tremaine, S., 1980, ApJ, 241, 425
\reference{G06} Gould, A. et al., 2006, ApJ, 644, L37
\reference{G10} Gould, A. et al., 2010, ApJ, 720, 1073
\reference{H09} Hansen, B., 2009, ApJ, 703, 1131
\reference{And} Hansen, B. \& Murray, N., 2012, ApJ 751, 158
\reference{HP} Hasegawa, Y. \& Pudritz, R. E., 2011, MNRAS, 417, 1236
\reference{HM05} Holman, M. J. \& Murray, N., 2005, Science, 307, 1288
\reference{H10} Holman, M. J. et al., 2010, Science, 330, 51
\reference{How} Howard, A. et al., 2010, Science, 330, 653
\reference{H11} Howard, A. et al., 2012, ApJS, 201, 15
\reference{IL08} Ida, S. \& Lin, D. N. C., 2008, ApJ, 685, 584
\reference{IL10} Ida, S. \& Lin, D. N. C., 2010, ApJ, 719, 810
\reference{JGM} Johnson, E. T., Goodman, J. \& Menou, K., 2006, ApJ, 647, 1413
\reference{K08} Kalas, P. et al., 2008, Science, 322, 1345
\reference{KI98} Kokubo, E. \& Ida, S., 1998, Icarus, 131, 171
\reference{KL} Kretke, K. \& Lin, D. N. C., 2012, ApJ, 755, 74
\reference{L08} Laskar, J., 2008, Icarus, 196, 1
\reference{LSA} Laughlin, G., Steinacker, A. \& Adams, F. C., 2004, ApJ, 608, 489
\reference{LA} Levison, H. F. \& Agnor, C. A., 2003, AJ, 125, 2692
\reference{LBR} Lin, D. N. C., Bodenheimer, P. \& Richardson, D. C., Nature, 380, 606
\reference{L11} Lissauer, J. J. et al., 2011a, Nature, 470, 53
\reference{L11b} Lissauer, J. J. et al., 2011b, ApJS, 197, 8
\reference{L12} Lissauer, J. J. et al., 2012, ApJ, 750, 112
\reference{LW11} Lithwick, Y. \& Wu, Y., 2011, ApJ, 739, 31
\reference{L07} Lovis, C. et al., 2006, Nature, 441, 305
\reference{M97} Marcy, G. W. et al., 1997, ApJ, 481, 926
\reference{M08} Marois, C. et al., 2008, Science, 322, 1348
\reference{M10} Marois, C. et al., 2010, Nature, 468, 1080
\reference{MMCF} Masset, F., Morbidelli, A., Crida, A. \& Ferreira, J., 2006, ApJ, 642, 478
\reference{MIN} Matsumura, S., Ida, S. \& Nagasawa, M., 2012, arXiv:1209.1320
\reference{MQ} Mayor, M. \& Queloz, D., 1995, Nature, 378, 355
\reference{M09} Mayor, M. et al., 2009, A\&A, 507, 487
\reference{M11} Mayor, M. et al., 2011, arXiv:1109.2497
\reference{Maz} Mazeh, T. et al., 2013, arXiv:1301.5499
\reference{Mc04} McArthur et al., 2004, ApJ, 614, L81
\reference{MJ11} Morton, T. D. \& Johnson, J. A., 2011, ApJ, 738, 170
\reference{MF} Moorhead, A. V. et al., 2011, ApJS, 197, 1
\reference{NIL} Nagasawa, M., Lin, D. N. C. \& Ida, S., 2003, ApJ, 586, 1374
\reference{NI} Nagasawa, M. \& Ida, S., 2011, ApJ, 742, 72
\reference{NN} Naoz, S., et al., 2011, Nature, 473, 187
\reference{PM} Paardekooper, S. J. \& Mellema, G., 2006, A\&A, 459, L17
\reference{PBCK} Paardekooper, S. J., Baruteau, C., Crida, A. \& Kley, W., 2010, MNRAS, 401, 1950
\reference{RF96} Rasio, F. A. \& Ford, E. B., 1996, Science, 274, 954
\reference{R12} Rein, H., 2012, MNRAS, 427, L21
\reference{Riv05} Rivera, E., et al., 2005, ApJ 634, 625
\reference{Q09} Queloz, D. et al., 2009, A\&A, 506, 303
\reference{S12} Santerne, A. et al., 2012, A\&A, 545, 76
\reference{S04} Santos, N. C. et al., 2004, A\&A, 426, L19
\reference{SL09} Smith, A. W. \& Lissauer, J. J., 2009, Icarus, 381
\reference{ST12} Steffen, J. et al., 2012, ApJ, 756, 186
\reference{ST13} Steffen, J., 2013, arXiv:1301.2394
\reference{S10} Sumi, T. et al., 2010, ApJ, 710, 1641
\reference{S11} Sumi, T. et al., 2011, Nature, 473, 349
\reference{T03} Terquem, C., 2003, MNRAS, 341, 1157
\reference{TP07} Terquem, C. \& Papaloizou, J. C. B., 2007, ApJ, 654, 1110
\reference{TNL} Thommes, E., Nagasawa, M. \& Lin, D. N. C., 2008, ApJ, 676, 728
\reference{T12} Tuomi, M. et al., 2012, arXiv:1211.1617
\reference{Ud} Udalski, A. et al., 2005, ApJ, 625, L109
\reference{VA} Veras, D. \& Armitage, P. J., 2005, ApJ, 620, L111
\reference{VFP} Veras, D., Ford, E. B. \& Payne, M. J., 2011, ApJ, 727, 74
\reference{W81} Ward, W. R., 1981, Icarus, 47, 234
\reference{WFAJ} Winn, J. N., et al., 2010, ApJ, 718, L145
\reference{WF} Wolszczan, A. \& Frail, D. A., 1992, Nature, 355, 145
\reference{WL1} Wu, Y. \& Lithwick, Y., 2012b, ApJ, 756, L11
\reference{WL2} Wu, Y. \& Lithwick, Y., 2012a, arXiv:1210.7810
\reference{WM} Wu, Y. \& Murray, N., 2003, ApJ, 589,605
\reference{Y12} Youdin, A., 2011, ApJ, 742, 38
\end{references}
\end{document}